\documentclass[journal=jctcce,manuscript=article]{achemso}
\usepackage[version=3]{mhchem} % Formula subscripts using \ce{}
\usepackage[dvipsnames]{xcolor}
\usepackage{graphicx,subcaption}
\usepackage{amssymb}
\usepackage{braket}
\usepackage{bm}
\usepackage{derivative}
\usepackage{hyperref}
\usepackage{siunitx}
\usepackage{booktabs}
 \DeclareSIUnit\angstrom{\text {Å}}
\author{Vishikh Athavale}
\email{vishikh@lanl.gov}
\author{Nikita Fedik}
\author{William Colglazier}
\altaffiliation{Texas A$\&$M University, College Station, Texas 77840, United States}

\author{Anders M. N. Niklasson}
\author{Maksim Kulichenko}
\email{maxim@lanl.gov}
\author{Sergei Tretiak}
\affiliation[Unknown University]
{Theoretical Division, Los Alamos National Laboratory, Los Alamos, New Mexico 87545, United States} 
\alsoaffiliation[Second University]
{Center for Nonlinear Studies and Center for Integrated Nanotechnologies, Los Alamos National Laboratory, Los Alamos, New Mexico 87545, United States}
\title{PySEQM 2.0: Accelerated Semiempirical Excited State Calculations on Graphical Processing Units}
\begin{document}

% \begin{tocentry}

% \end{tocentry}

\begin{abstract}
We report the implementation of electronic excited states for semi-empirical quantum chemical methods at the configuration interaction singles (CIS) and time-dependent Hartree Fock (TDHF) level of theory in the PySEQM software. Built on PyTorch, this implementation leverages GPU acceleration to significantly speed up molecular property calculations. Benchmark tests demonstrate that our approach can compute excited states for molecules with nearly a thousand atoms in under a minute. Additionally, the implementation also includes a machine learning interface to enable parameters' re-optimization and neural network training for future machine learning applications for excited state dynamics.
\end{abstract}

\section{Introduction}
%[General Intro to excited states + ref to Fig 1]
Modeling photophysical and photochemical phenomena in molecular systems requires accurate and fast computation of electronic excited states. However, excited state calculations pose a significant challenge in quantum chemistry due to their theoretical complexity and high computational overhead. 
For small molecular systems, sophisticated quantum chemical methods such as equation-of-motion and linear-response coupled-cluster methods\cite{bartlett:1993:eomccsd} provide high accuracy for studying electronic excited states\cite{szalay:2020:NewBenchmarkSet:J.Chem.TheoryComput.}.
Algebraic diagrammatic construction (ADC) methods\cite{schirmer:1982:ADC:Phys.Rev.A,schirmer:1995:ADC2:J.Phys.B} can offer a balanced alternative to coupled cluster techniques, delivering robust excited state descriptions at a moderate computational cost\cite{wormit:2015:ADC:WIREsComput.Mol.Sci.}. 
In cases where static correlations play a critical role in the molecule's electronic structure, multi-configurational approaches such as complete active space self-consistent field (CASSCF)\cite{sigbahn:1980:casscf:Chem.Phys.} and its perturbative extension (CASPT2)\cite{wolinski:1990:caspt2:J.Phys.Chem.,roos:1992:caspt2_full:J.Chem.Phys.} are very useful for small to medium sized systems. 
For molecules containing hundreds of atoms, time-dependent density functional theory (TDDFT)\cite{gross:1984:tddft_prl,furche:2001:tddft_jcp,casida:tddft:book} is routinely used to calculate excited states with good accuracy\cite{yihan:2020:tddft_benchmark,jctc:2009:tddft_benchmark,truhlar:2012:tddft_benchmark,furche:2011:tddft_benchmak:jctc,headgordon:2011:tddft_benchmark,mark_gordon:2012:tddft_benchmark}.  Notably, $\Delta$SCF methods can be used to calculate targeted excited states through an orbital-optimization-based approach that have shown particular strength in describing charge-transfer and Rydberg excitations, and excited states with multiple-excitation character\cite{barca_simple_2018,hait_highly_2020,carter-fenk_state-targeted_2020,kunze_pcm-roks_2021}. While the aforementioned methods and their variants are some of the most commonly used excited state methods today, the system sizes for which chemical processes occurring on the excited states can be investigated is limited by the computational costs.
%, despite its known limitation in treating Rydberg states\cite{handy:1998:tddft_rydberg,casida:1998:tddft_rydberg} and charge-transfer states\cite{tozer:2003:ct_failure_tddft,headgordon:2003:ct_failure_tddft}.

% \begin{figure}[hptb]
%     \centering

%     \includegraphics[scale=.7]{figures/Slide1.PNG}
%     \caption{A. Light-matter interaction drive a plethora of  chemical and physical phenomena of interest. Here, we focus on single excitation picture when one electron from occupied orbital is promoted to the virtual orbital. Consequently, in a dynamical picture, system can switch from one potential energy surface to another. In general, molecular excitations cause various transformations such as change of coloration, photodissociation or change transfer, name merely a few examples.  B.    
%     An overview and a concise roadmap of the capabilities of the PySEQM2.0 framework, leveraging GPU-support and semiempirical Hamiltonians for fast chemical simulations. Version 2.0 brings a new module to compute molecular excited states using Davidson algorithm. In addition, to CPU and standard GPU modes (one molecule per unit), we provide a batch mode handling multiple molecules on a single GPU in parallel. 
%    }
%     \label{fig:pyseqm_features}
% \end{figure}

Semiempirical quantum mechanics (SEQM) elevate the scale of molecular simulations up to thousands of atoms. By leveraging various simplification and approximations \cite{seqm_review:2016:chem_rev}, SEQM methods achieve a more favorable computational scaling with system size and a lower prefactor of time complexity compared to DFT and \textit{ab initio} methods. For example, time-dependent density functional tight binding (TD-DFTB)\cite{niehaus_tight-binding_2001,dominguez_extensions_2013, nishimoto_time-dependent_2015, kranz_time-dependent_2017} can drastically bring down the cost of excited state calculations. While the accuracy of TD-DFTB methods is generally lower compared to TDDFT, some extensions to TD-DFTB such as TD-DFTB3\cite{nishimoto_time-dependent_2015}, and LC-DFTB2\cite{lutsker_implementation_2015,nishimoto_time-dependent_2019} can match or even surpass the performance of conventional TDDFT\cite{fihey_performances_2019, bertoni_data-driven_2023}. Similarly, SEQM based on the neglect of diatomic differential overlap (NDDO) offers a more computationally efficient alternative to \textit{ab initio} methods. Excited states can be calculated from these Hamiltonians using time-dependentent Hatree Fock (TD-HF) theory and multi-reference configuration interaction (MRCI) methods\cite{koslowski_implementation_2003}. 
Improvements in the NDDO Hamiltonian itself, such as the OMx family of SEQM methods\cite{kolb_beyond_1993,weber_orthogonalization_2000}, can further enhance the description of excited states.\cite{silva-junior_benchmark_2010,dral_semiempirical_2016}
Ultimately, these SEQM methods lower the computational barrier for nonadiabatic dynamics\cite{liu_efficient_2018} to study complex excited state processes in large systems over extended time scales\cite{zobel_quest_2021, westermayr_machine_2019, akimov_excited_2021, patel_mixed_2020, pal_nonadiabatic_2016}. %XXX 

With rapid progress in graphical processing unit (GPU) and specialized AI-hardware, quantum chemistry has further pushed the limits of the systems it can deal with \cite{TMartinez08,TGermann09,TMartinez09a,TMartinez09b,SGoedecker09,JStone10,TMartinez11,JMaia12,MHacene12,MCawkwell12b,FLiu2015,WHuhn20,MGordon20,fales_nanoscale_2015,ZGuoqing20,JFinkelstein21,JFinkelstein21B}. In the context of excited states, it is feasible to calculate excited state energies as well as analytical gradients for solvated proteins with more than 4000 atoms at the TDDFT level\cite{kim_kohnsham_2023,kim_very-large-scale_2024}.
This progress has been driven in part by the rise of machine learning (ML), which has catalyzed the development of accessible frameworks for building GPU-accelerated software. For example, consider PyTorch--a user-friendly Python library, originally developed for dealing with ML problems. Using PyTorch, the cost-intensive linear algebra operations can be efficiently vectorized and run on GPUs. Leveraging this capability, the GPU-accelerated SEQM code PySEQM\cite{pyseqm:2020:jctc} was developed for efficient ground-state molecular dynamics simulations, while providing an easy interface for machine learning of molecular properties\cite{zhou_deep_2022}.
See below for an overview of all the features of PySEQM.

In this article, we present new capabilities in PySEQM with the implementation of excited state calculations using the time-dependent Hartee-Fock (TDHF) and configuration interactions singles (CIS) for SEQM methods. We test the scalability of our approach with increasing system size and demonstrate that our implementation is able to calculate twenty excited states for systems with a thousand atoms in under a minute on a single GPU. We expect this capability to be particularly important for applications in excited state molecular dynamics, enabling the study of large systems where \textit{ab initio} and DFT methods are computationally prohibitive.

\section{Overview of the capabilities of PySEQM}
PySEQM is a high-performance quantum chemistry software designed to exploit GPU acceleration for efficient electronic structure and molecular dynamics simulations. The ground state wave function is calculated using SEQM methods based on the NDDO approximation, such as MNDO\cite{dewar_ground_1977}, AM1\cite{dewar_development_1985}, PM3\cite{stewart_optimization_1989}, and PM6\cite{stewart_optimization_2007}.  By leveraging PyTorch’s automatic differentiation, PySEQM can compute energy gradients directly, enabling both geometry optimizations and molecular dynamics with minimal overhead. For dynamics simulations, PySEQM offers a flexible molecular dynamics (MD) driver that supports Born–Oppenheimer Molecular Dynamics (BOMD). Additionally, the extended-Lagrangian BOMD (XL-BOMD) scheme\cite{niklasson_extended_2008, niklasson_extended_2009, niklasson_density-matrix_2020, niklasson_extended_2021, kulichenko_semi-empirical_2023} is available, which improves long-term energy conservation and avoids the costly overhead and potential instabilities associated with insufficiently converged self-consistent field (SCF) calculations. These MD capabilities allow users to investigate ground-state chemical processes while harnessing powerful GPU resources. 

One of PySEQM's distinctive features is providing a framework for ML. Often ML models for molecular properties merely train on a large dataset to make predictions without revealing anything about the underlying physics.
However, with PySEQM, ML is used to optimize the underlying SEQM Hamiltonians without the loss of physical transparency. Such an approach is highly transferable and hence extensible to novel chemical systems. In fact, PySEQM has already been interfaced with the HIP-NN neural network model\cite{lubbers_hierarchical_2018}, which dynamically predicts semiempirical Hamiltonian parameters leading to improved molecular property calculations\cite{zhou_deep_2022}.

\begin{figure}[hptb]
    \centering
    \includegraphics[scale=.7]{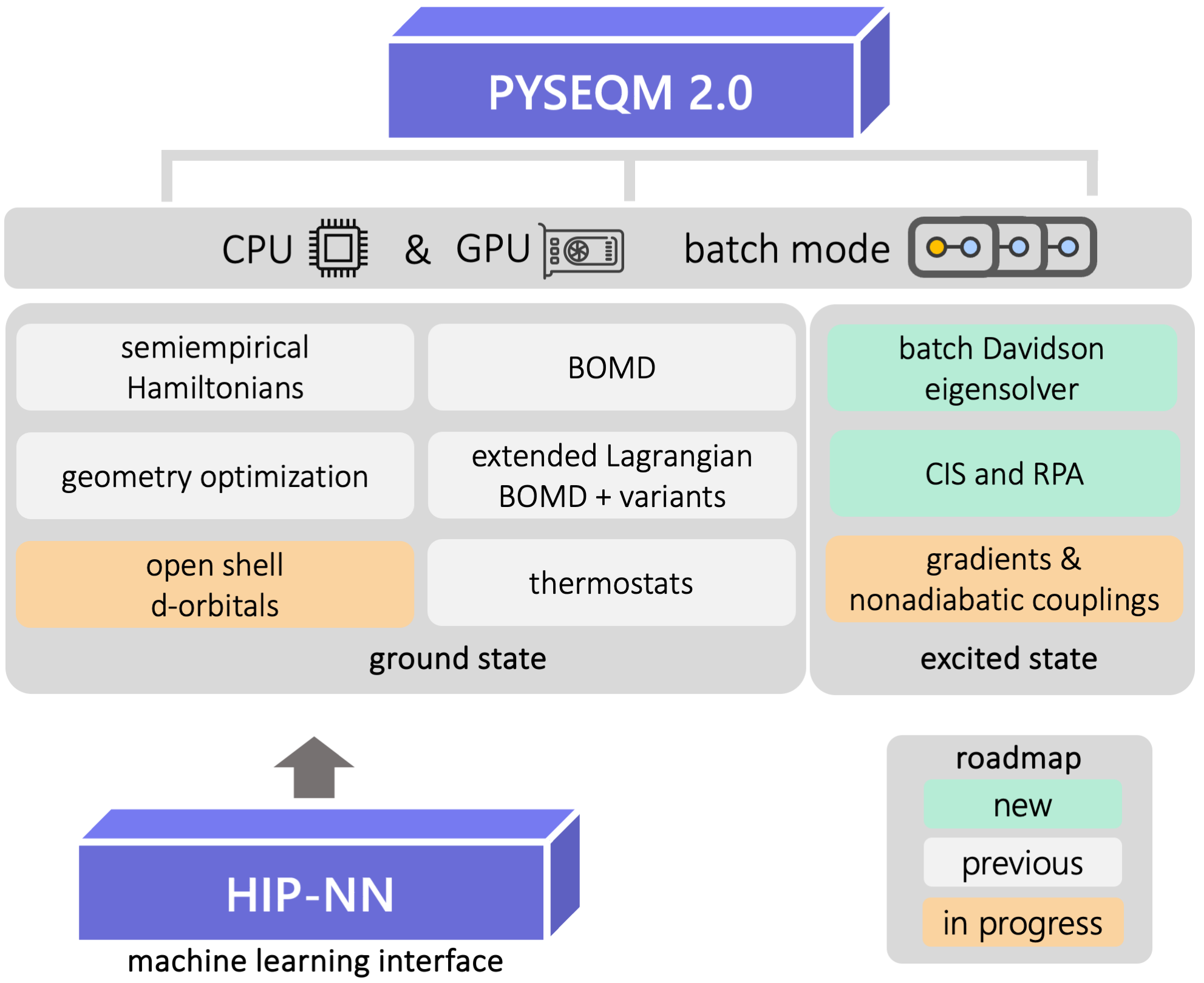}
    \caption{An overview of the capabilities of the PySEQM software, highlighting the available features, newly developed features, and features that are forthcoming.}
    \label{fig:pyseqm_features}
\end{figure}

Building on this foundation, we extend the software into excited-state calculations, offering an extremely fast implementation that runs on modern GPU hardware. PySEQM is under active development with work being done on adding additional features such as support for open-shell systems, and for atoms with d-type orbitals in their basis sets. We will also add support for excited state gradients and nonadiabatic couplings, enabling the study of large-scale photoactive materials in complex environments. Fig.~\ref{fig:pyseqm_features} provides an outline of the features available in PySEQM.
 
\section{Theory and implementation}
\subsection{Ground-state wavefunction calculation}
In PySEQM, the ground state wave function is calculated using SEQM methods based on the NDDO approximation, as explained above. These methods employ a minimal basis set of Slater-type orbitals (STOs), explicitly considering only the valence electrons for each element. All three- and four-center two-electron integrals are neglected, greatly reducing the computational cost. Two-center electron repulsion integrals that represent interactions between two atomic charge centers are retained and approximated using classical multipole interactions. Additionally, one-center integrals are incorporated through empirically derived parameters that are fitted to experimental data or high-level \textit{ab initio} reference calculations.

The molecular orbital coefficients $\bm{C}$ are obtained with the Self Consistent Field (SCF) approach by solving the secular equation
\begin{equation}
    \bm F \bm C = \bm{C\epsilon},
\end{equation}
where $\bm F$ is the Fock matrix, and $\bm \epsilon$ is the diagonal matrix of orbital energies. The overlap integral matrix under the NDDO approximation becomes the identity matrix, yielding the above secular equation.  
Assuming a restricted closed-shell ground state, the wave function $\ket{\Psi_0}$ can be formed as a single Slater determinant comprising $N$ molecular orbitals of lowest energy for a system with $2N$ electrons (each molecular orbital is doubly occupied). 

In the following sections, we briefly outline the theoretical framework of TDHF and CIS, highlighting the key equations used to calculate excited states. We then describe the implementation of these methods in PySEQM, focusing on the key considerations required for efficient GPU utilization.

\subsection{TDHF and CIS equations}
Two common single excitation methods used to calculate excited states are TDHF and CIS\cite{dreuw_single-reference_2005}. TDHF applied for the calculation of electronic excited states is also known as random phase approximation (RPA). Electronic excited energies $\omega$ are obtained by solving the RPA eigenvalue equation
\begin{equation}
    \begin{bmatrix}
    \bm{A} & \bm B \\
    \bm{B} & \bm{A}
    \end{bmatrix} 
    \begin{bmatrix}
        \bm X \\ \bm Y
    \end{bmatrix} 
    =\omega
    \begin{bmatrix}
    \bm{I} & \bm 0 \\
    \bm{0} & -\bm{I}
    \end{bmatrix} 
    \begin{bmatrix}
        \bm X \\ \bm Y
    \end{bmatrix}.\label{eq:rpa_eqn}
\end{equation}
The individual elements of the $\bm{A}$ and $\bm{B}$ matrices in the molecular orbital representation are given by
\begin{align}
    A_{ia,jb} &= (\epsilon_a-\epsilon_i)\delta_{ij}\delta_{ab} + (ia|jb) - (ij|ab),\label{eq:A_matrix_elem}\\
    B_{ia,jb} &= (ia|jb) - (ib|ja).
\end{align}
Here and below, we use the indices $i, j$ for occupied molecular orbitals and  $a, b$  for virtual molecular orbitals; spin indices have been omitted. In Eq.~\ref{eq:A_matrix_elem} above,
$\epsilon_i$, $\epsilon_j$ represent the orbital energies of the molecular orbitals $\phi_i$, $\phi_a$ respectively, and $(ia|jb)$ represents the two-electron integral
\begin{equation}
    (ia|jb)=\int \odif{r,r'}\frac{\phi_i(r)\phi_a(r)\phi_j(r')\phi_b(r')}{|r-r'|}.
\end{equation}

Another approach to obtaining excited states starting from the Hartree-Fock reference wave function is through the CIS ansatz. The CIS wave function $\ket{\Psi_{\mathrm{CIS}}}$ is constructed by allowing all single excitations from the occupied orbitals to the virtual orbitals
\begin{equation}
    \ket{\Psi_{\mathrm{CIS}}}=
    \sum_{ia}X_{ia}\ket{\Phi_i^a}.
\end{equation}
Here $\ket{\Phi_i^a}$ denotes the Slater determinant formed by replacing the occupied orbital $i$ with the virtual orbital $a$ in the reference HF determinant $\ket{\Psi_0}$, and $X_{ia}$ is the amplitude corresponding to $\ket{\Phi_i^a}$. 
The CIS amplitudes and excitation energies $\omega$ are obtained by solving the eigenvalue equation
\begin{equation}
    \bm{AX}=\omega\bm{ X}.\label{eq:cis_eqn}
\end{equation}
The CIS eigenvalue equation can also be obtained by setting $\bm B=0$ in the RPA eigenvalue equation (eq.~\ref{eq:rpa_eqn}). Such an approximation to RPA is known as the Tamm-Dancoff approximation (TDA).

\subsection{GPU Implementation}
TDHF involves solving the non-Hermitian eigenvalue equation Eq.~\ref{eq:rpa_eqn}, which is more expensive than the simple Hermitian eigenvalue equation for CIS (eq.~\ref{eq:cis_eqn}). In Eq.~\ref{eq:rpa_eqn}, the matrices $\bm A$ and $\bm B$ are of the size $(N_{\mathrm{occ}}N_{\mathrm{vir}})^2$, where $N_{\mathrm{occ}}$ is the number of occupied orbitals and $N_{\mathrm{vir}}$ is the number of virtual orbitals. In practice, these matrices are never explicitly built to be diagonalized because of the high computational cost and memory requirements. Instead, iterative diagonalization techniques such as the Davidson algorithm\cite{davidson_iterative_1975,liu_simultaneous_1978} are used to calculate the lowest few eigenstates of Eq.~\ref{eq:rpa_eqn} and \ref{eq:cis_eqn}. Taking the example of CIS eigenvalue equation (Eq.~\ref{eq:cis_eqn}), the Davidson algorithm involves building $\bm A$ in a smaller subspace and only requires matrix-vector products between $\bm A$ and the set of subspace basis vectors $\bm{V}$. For a basis vector $\bm V^I$, the matrix-vector product is 
\begin{align}
\label{eq:cis_matrix_vector}
    (\bm{AV}^I)_{ia} &= \sum_{jb}A_{ia,jb}V^I_{jb}\nonumber\\
    &= (\epsilon_a-\epsilon_i)V^I_{ia} + \sum_{jb}\left[(ia|jb) - (ij|ab)\right]V^I_{jb}\nonumber\\
    &= (\epsilon_a-\epsilon_i)V^I_{ia} + \sum_{\mu\nu}C_{\mu i}\Tilde{F}_{\mu\nu}C_{\nu a},
\end{align}
where $\Tilde{\bm F}$ is defined as
\begin{equation}
    \Tilde{F}_{\mu\nu}\equiv\sum_{\lambda\sigma}\left[(\mu\nu|\lambda\sigma)-(\mu\lambda|\nu\sigma)\right]R_{\lambda\sigma}^I, \label{eq:ftilde}
\end{equation}
and the density matrix $\bm{R}^I$ is defined as $R^I_{\lambda\sigma}\equiv\sum_{jb}C_{\lambda j}C_{\sigma b}V^I_{jb}.$
In the above equations, ${\mu}$, ${\lambda}$, ${\nu}$, and ${\sigma}$ are atomic orbital indices.

For NDDO methods, four- and three-center two-electron integrals are neglected, and only certain two-center integrals need to be calculated (while one-center electron repulsion integrals are tabulated parameters). The two-center two-electron integrals are generally calculated before the SCF cycle and then stored in memory. The construction of $\Tilde{\bm F}$ involves the contraction of these two-electron integrals with the nonsymmetric density matrix $\bm R^I$. 
This differs from the ground state Fock build, where the density matrix is symmetric. Hence, the Fock build routines in PySEQM required a slight modification to account for the nonsymmetric density matrix. As can be seen in Eq.~\ref{eq:ftilde}, $\Tilde{\bm F}$ consists of the Coulomb  ($\sum_{\lambda\sigma}(\mu\nu|\lambda\sigma)R_{\lambda\sigma}^I$) and the exchange matrix elements ($\sum_{\lambda\sigma}(\mu\lambda|\nu\sigma)R_{\lambda\sigma}^I$). In the 
 NDDO approximation, the Coulomb matrix elements contribute only in the diagonal blocks of  $\Tilde{\bm F}$. Because of the symmetry present in the two electron integrals, a symmetrized form of $\bm R^I$ is used while contracting with the two electron integrals in the AO basis to obtain the Coulomb matrix elements. This is analogous to the ground state Fock build. However, for calculating the exchange matrix elements, we have to call the integral contraction routine twice: once with the symmetrized and once with anti-symmetrized form of $\bm R^I$.

PySEQM supports a ``batch" mode that processes multiple molecular systems concurrently. This design ensures maximal utilization of the GPUs and reflects a common approach in machine learning where batched operations are used to optimize training and testing workflows. 
Additionally, several nonadiabatic dynamics methods (e.g. surface-hopping) require an ensemble of trajectories, and a batched implementation directly facilitates this requirement. 
Here we present a step-by-step outline of the ``batched" Davidson algorithm for the calculation of CIS excited states for a set of molecules as implemented in PySEQM.
We retain the standard Davidson method while introducing batch-mode modifications to handle multiple molecular calculations in parallel. Note that in the steps below, boldface symbols (e.g. $\bm{V,A}$) denote three-dimensional tensors rather than the usual two-dimensional matrices, with the extra index running over the batch of molecules.

\begin{figure}[hptb]
    \centering
    \includegraphics[scale=1]{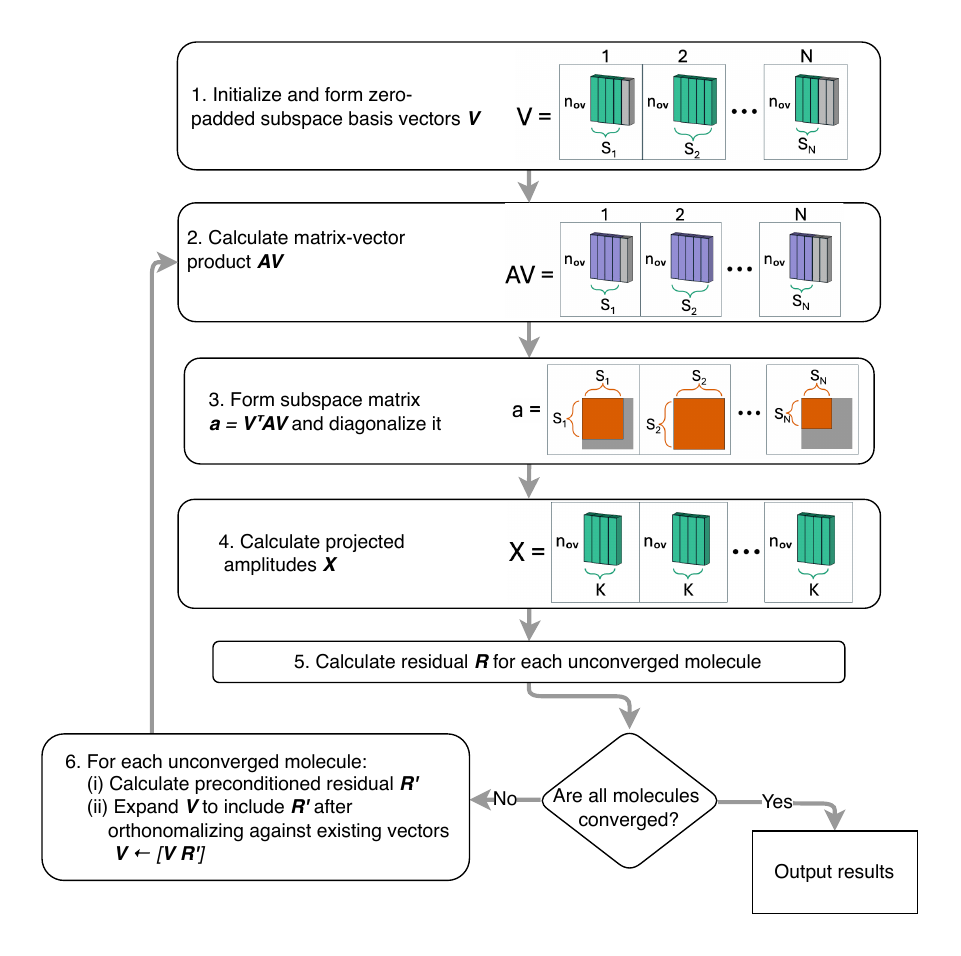}
    \caption{Flowchart of the batched Davidson algorithm in PySEQM for computing CIS excitation energies across multiple molecules in parallel}
    \label{fig:Davidson_flowchart}
\end{figure}

The objective is to obtain the lowest $K$ eigenstates for $N$ molecules by solving  Eq.~\ref{eq:cis_eqn} for each molecule. 
\begin{enumerate}
    \item Begin with a set of orthonormal subspace basis vectors for each molecule and pack these vectors into a 3-D tensor $\bm V$. All molecules are initially flagged as unconverged. 
    \item Compute batched matrix-vector products between the CIS Hamiltonian $\bm A$ and the subspace basis vectors $\bm V$ (Eq.~\ref{eq:cis_matrix_vector}). Before doing so, prepare the tensor $\bm V$. 
    As the Davidson iteration proceeds, each molecule $i$ will have its own subspace size $S_i$ (see Fig.~\ref{fig:Davidson_flowchart}). Let $S_{\max} = \max\{S_1, S_2, \dots, S_N\}$. For any molecule whose current subspace $S_i$ is smaller than $S_{\max}$, append $S_{\max}-S_i$ zero columns to its slice of $\bm V$.  These zero columns carry no physical information but enforce a uniform slice width of $S_{\max}$ across the batch. This guarantees that the tensor shapes match and allows the GPU to perform one highly efficient, fused batched multiplication.
    %If one molecule’s subspace has fewer vectors than another’s, pad the vector $\bm{V}$ with zero columns so every slice of $\bm V$ has the same width, called $S_{\text{max}}$. This zero-padding is necessary to have uniformly shaped tensors such that batch matrix multiplication can be efficiently parallelized by the GPU kernel.
    \item Calculate the projected Hamiltonian $\bm a = \bm{V}^{\dagger}\bm{A}\bm V$ for each unconverged molecule and diagonalize $\bm a$. Discard the eigenstates that arise from the zero-padding in $\bm a$. Following this, extract the lowest $K$ eigenvalues $\bm\omega$ and eigenvectors $\bm x$.
    \item  Compute $K$ projected amplitudes $\bm X$ for each molecule from the eigenvectors using $\bm{X=Vx}$.
    \item Calculate the residual vectors $\bm r$ for every candidate state ($K$ per molecule) using $\bm{r = AX - \omega X}$.
    Molecules for whom the norm of every residual vector is below tolerance are flagged as converged.
    \item  For the unconverged molecules, apply standard Davidson diagonal preconditioning to only those residual vectors that have not yet met the convergence threshold, leaving already-converged states untouched. Orthogonormalize pre-conditioned residuals against the existing basis vectors in $\bm V$ and append them to $\bm V$. Return to Step 2 and repeat until all eigenstates meet the convergence condition.
\end{enumerate}

%In general, molecules reach convergence for different eigenstates at different rates. For those molecules where some of their eignestates converge early, the number of new subspace basis vectors that are added at the end of an iteration can be lesser than those for other molecules. In this case, for molecules with a subspace size smaller than the maximum among all the molecules, $S_{\mathrm{max}}$, the extra subspace expansion vectors are treated as zeros in $\bm V$.  
%Then, the subspace Hamiltonian $\bm a$ is built and diagonalized. The eigenstates with zero eigenvalue in $\bm a$ that occur because of zero-padding in $\bm V$ are first discarded, and then the lowest $K$ eigenstates for each molecule are retained. We compute residuals for all eigenstates, apply preconditioning to unconverged states, and orthogonalize the resulting vectors against $\bm V$.
%This process continues until all required eigenstates of all molecules meet the specified convergence threshold.

The above algorithm has been schematically represented in the flowchart in Fig.~\ref{fig:Davidson_flowchart}, which makes the dimensions of the tensors involved clear from their diagrams. We expect the process to be well balanced for molecules with similar geometry, e.g. when simulating an ensemble of molecules to sample a region of the potential energy surface. For pathological cases where one molecule in the batch takes far greater time to converge than all the other molecules, our implementation may require a redesign.

In our implementation of the Davidson algorithm for CIS and RPA, the maximum size of the Krylov subspace is determined by hardware constraints, i.e., the maximum subspace size is calculated as a fraction of the available free space in GPU memory. During the Davidson iterations, the formation of matrix vector products (Eq.~\ref{eq:cis_matrix_vector}) is another memory intensive step. When the available free memory is not sufficient to calculate matrix-vector products, especially when the number of vectors is large, then the matrix-vector products are calculated in smaller batches that fit in the available memory. This scheme has been very effective such that we are able to calculate up to 100 excited states of a 1000-atom molecule without any out-of-memory issues on the GPU, as we show in the section below. 

\section{Results and discussion}
In this section, we illustrate the performance of excited state calculations in PySEQM on GPU hardware. Our implementation supports both TDHF as well as CIS. In \textit{ab initio} theory, TDHF offers no clear practical advantage over CIS to justify its increased cost (both methods generally overestimate excitation energies). Consequently, we focus our performance discussion on CIS calculations only. 

\begin{figure}[hptb]
    \centering
    \includegraphics[scale=0.8]{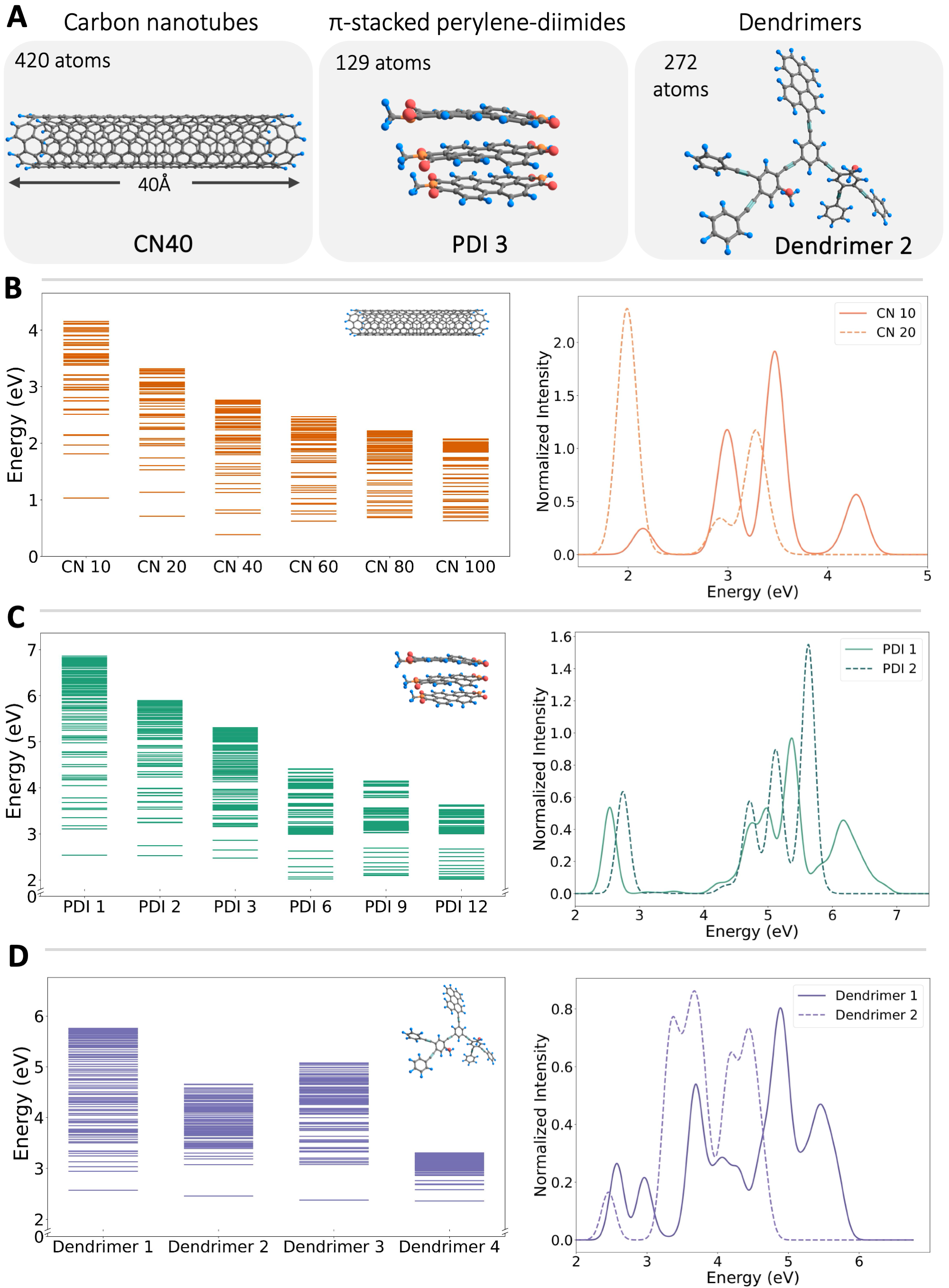}
    \caption{(A) Structures of CN-40, PDI-3, and Dendrimer-2, which serve as representative molecules from the three molecular test sets examined in the present study. Images of all remaining molecules can be found in the Supporting Information. 
    (B-D) Visualization of the excited-state manifolds for (B) Carbon nanotubes, (C) Perylene diimide stacks, and (D) Dendrimers.
    In each panel, the left subfigure shows the density of states of the computed CIS excited states for all molecules in that series, and the right subfigure depicts the simulated absorption spectra for the two smallest molecules from that family. }
    \label{fig:molecules}
\end{figure}

The correctness of our implementation of CIS and TDHF in PySEQM was verified by comparing the excitation energies obtained with ORCA 5.0.4\cite{neese_software_2022}. For a set of 28 small organic molecules benchmarked by Thiel and coworkers\cite{silva-junior_benchmark_2010}, the mean absolute error in excitation energies calculated by PySEQM versus ORCA for the lowest 5 CIS excited states of each molecule was \SI{0.18}{\milli\electronvolt} and RPA excited states was \SI{0.19}{\milli\electronvolt} 
%(see Supporting Information for more details)
. 
We then benchmarked the performance of CIS using three sets of molecular systems with diverse structural features and sizes: 
\begin{enumerate}
    \item Single-walled carbon nanotubes (CNs) with chirality (6, 6) of increasing length from \SI{10}{\angstrom} to \SI{100}{\angstrom}, labeled CN-$n$, where $n= 10, 20, 40, 60, 80, 100,$ and $n$ denotes the length of the carbon nanotube.
    \item A set of four branched conjugated dendrimers (Dendrimer-1 through -4). Dendrimer-1 and Dendrimer-3 are the unsymmetrical phenylacetylene dendrimers referred to as 2G1(meta)Per and 2G2(meta)Per in ref.~\citenum{dendrimer:2006:photoresearch}, respectively. Dendrimer-2 and Dendrimer-4 are phenylene ethynelene dendrimers known as half-nanostar and nanostar\cite{shortreed_directed_1997,palma_electronic_2010}, respectively.
    \item $\pi$-stacked perylene-diimide (PDI) derivatives. These systems are labeled PDI-$n$, and contain $n$ stacked monomer units, where $n=$ 1, 2, 3, 6, 9, and 12. 
\end{enumerate}
These representative systems have received a great deal of interest because of their opto-electronic properties and their applications in light-harvesting devices\cite{ackermann_biosensing_2022, wieland_carbon_2021, zacheo_efficient_2024, mann_electrically_2007,astruc_dendrimers_2010,kim_applications_1998,chaudhuri_enhancing_2011,zhou_magic-angle_2018,huang_perylene-34910-tetracarboxylic_2011,may_relationship_2011}, and hence make suitable test-cases for benchmarking. 
Fig.~\ref{fig:molecules}A presents one of the molecules in each of the three molecular sets. The images of all the molecules used in this study can be found in the Supporting Information.
Table~\ref{tab:no_of_atoms} summarizes the number of atoms in each of these molecules.  

\begin{table}[h!]
\centering
\begin{tabular}{@{}lr@{}}
\toprule
Molecule    & No. of atoms \\ \midrule
CN-10       & 132          \\
CN-20       & 228          \\
CN-40       & 420          \\
CN-60       & 612          \\
CN-80       & 804          \\
CN-100      & 996          \\ \midrule
Dendrimer-1 & 124          \\
Dendrimer-2 & 272          \\
Dendrimer-3 & 428          \\
Dendrimer-4 & 884          \\ \midrule
PDI-1       & 43           \\
PDI-2       & 86           \\
PDI-3       & 129          \\
PDI-6       & 258          \\
PDI-9       & 387          \\
PDI-12      & 516          \\ \bottomrule
\end{tabular}
% \begin{tabular}{|c|c|l|c|l|c|}
% \hline
% \textbf{Carbon Nanotubes} & \textbf{Atoms} & \textbf{Dendrimers} & \textbf{Atoms} & \textbf{PDIs} & \textbf{Atoms} \\
% \hline
% CN 10 & 132 & Dendrimer 1 & 124 & PDI 1 & 43 \\
% CN 20 & 228 & Dendrimer 2 & 272 & PDI 2 & 86 \\
% CN 40 & 420 & Dendrimer 3 & 428 & PDI 3 & 129 \\
% CN 60 & 612 & Dendrimer 4 & 884 & PDI 6 & 258 \\
% CN 80 & 804 &  &  & PDI 9 & 387 \\
% CN 100 & 996 &  &  & PDI 12 & 516 \\
% \hline
% \end{tabular}
\caption{Number of atoms in the molecules in each of the three molecular sets benchmarked in this study}
\label{tab:no_of_atoms}
\end{table}

The Austin Model 1 (AM1)\cite{dewar_development_1985} semiempirical Hamiltonian was used for all the calculations performed.
%, since it has been previously applied to model the excited state dynamics of extended conjugated molecular systems\cite{xxx}. 
GPU calculations were performed on NVIDIA Tesla A100 GPUs with 80 GB of RAM, and CPU computations were run on Intel Xeon Platinum 8480L hardware.

The excited state manifold for the three molecular systems investigated is visualized in the energy level digaram shown in Fig.~\ref{fig:molecules}B-D.
For each system, we calculated the lowest 100 CIS excited states.
As the molecular size increases, the density of states generally increases, leading to a narrower span of excitation energies within the same number of excited states. For instance, the 100 lowest excited states of PDI-1 span \qtyrange{2.54}{6.86}{eV}, whereas for PDI-12 this range narrows down to \qtyrange{2.02}{3.63}{eV}.
Increased conjugation and possible delocalization of excited states with increasing molecular size can have a pronounced effect on the excited-state manifold\cite{mukazhanova_impact_2023} and, consequently, the optoelectronic properties of these molecules. 
Fig~\ref{fig:molecules}B-D also presents the calculated absorption spectra for two representative molecules in each molecular set. These spectra were obtained by first calculating the transition dipole moments and oscillator strengths for each excited state, followed by a Gaussian broadening (with width = \SI{0.1}{\electronvolt}) applied to produce a continuous absorption profile. The resulting spectra would constitute the first step in guiding nonadiabatic dynamics simulations as it indicates the relative intensities and energies of optically accessible excited states. 

\begin{figure}[hptb]
    \centering
    \includegraphics[scale=0.8]{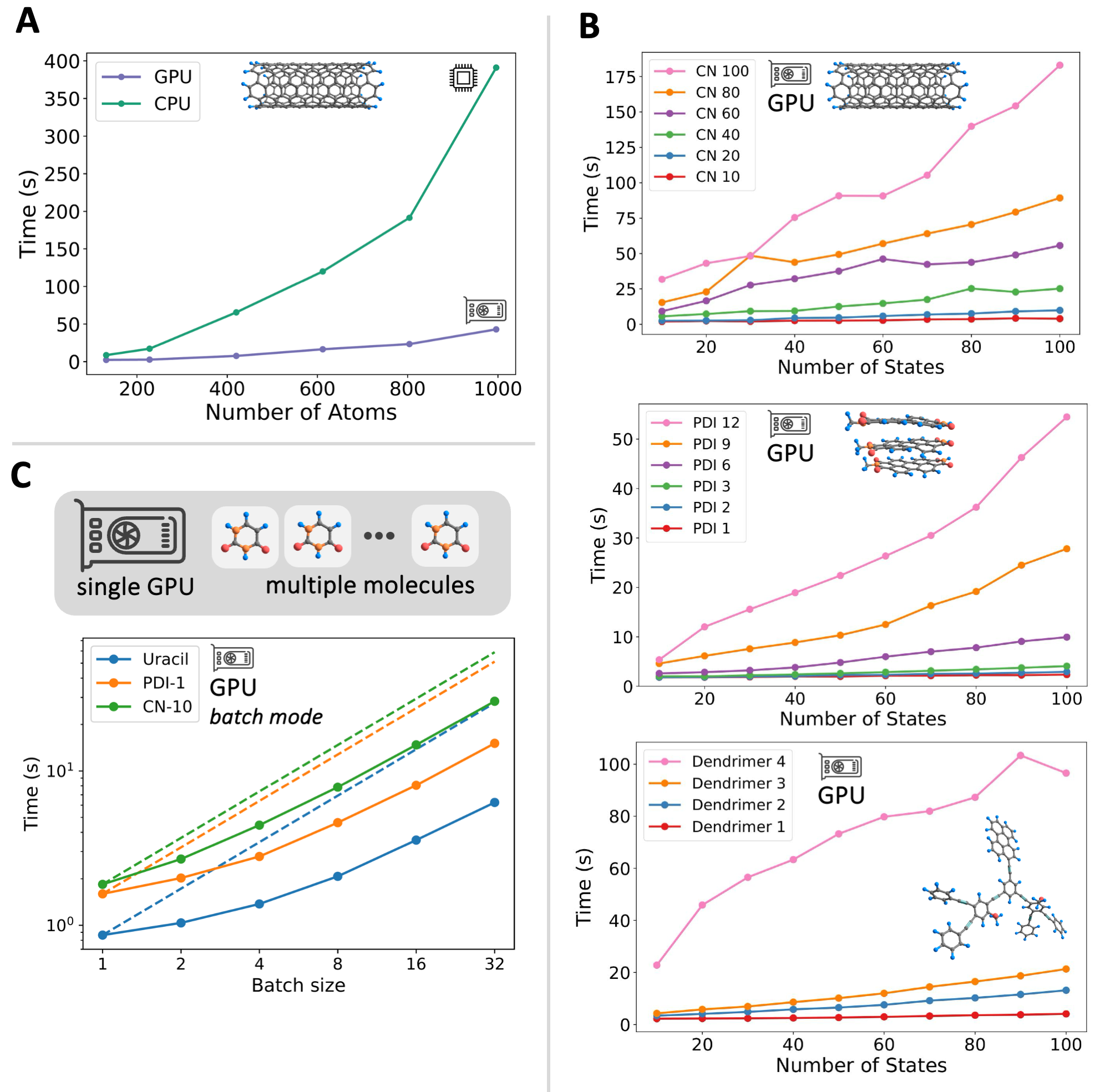}
    \caption{(A)Total computation time for calculating 20 CIS excited states in six carbon nanotube systems of increasing size, plotted against the number of atoms. Both CPU- and GPU-based runs in PySEQM are shown, highlighting the $\mathcal{O}(N^3)$ scaling typical of semiempirical CIS. GPU calculations deliver more than $8\times$ speed-up over CPU, yet even CPU-only runs remain reasonably fast.
    (B) Total computation time as a function of the number of CIS excited states shown for three molecular test sets: carbon nanotubes (CNs, top), perylene diimide (PDI, mid) stacks and dendrimers (bottom). Smaller systems (fewer than $\sim 200$ atoms) show nearly flat timing due to low GPU utilization, whereas medium-sized systems (200–500 atoms) scale more predictably with the number of states. Larger systems (over $\sim 500$ atoms) may trigger Krylov subspace collapses when GPU memory is exceeded, leading to non-monotonic timing increases. Despite occasional restarts, PySEQM computes up to 100 states efficiently, highlighting its robustness across varied molecular sizes.
    (C)Total time for running ``batched'' CIS excited state calculations on molecules uracil (12 atoms, shown in blue), PDI-1 (43 atoms, shown in orange) and CN-10 (132 atoms, shown in green) at various input geometries in increasing batch sizes. The corresponding dashed lines represent the predicted runtime if the computation time is scaled linearly with the batch size (extrapolated from single-molecule runtime). The deviation of the solid line from the dashed line represents the computational savings achieved by batching multiple calculations instead of running them individually.}
    \label{fig:timings}
\end{figure}

While the excited-state manifold and corresponding absorption spectra are insightful for nonadiabatic studies, the primary goal of this work is to assess the computational efficiency of GPU-accelerated excited-state calculations in PySEQM. To that end, Fig.~\ref{fig:timings}A shows the total time for calculating excited states for all the carbon nanotube molecules as a function of the number of atoms for a fixed number of states (in this case, 20). 
%In contrast to Fig.~\ref{fig:time_vs_states}, which illustrates how runtime varies with the number of excited states, Fig.~\ref{fig:num_atoms} highlights the scaling behavior for a fixed number of states (in this case, 20).
Formally, the time complexity for CIS calculated with the Davidson-like algorithm goes as $\mathcal{O}(N^4)$, where $N$ is the number of basis functions, which is proportional to the number of atoms. This is the same time complexity for Hartree-Fock and arises because of the calculation of the four-index two-electron integrals.
However, for NDDO-based SEQM methods, the approximations involved in the calculations of two-electron integrals (as discussed previously) reduce the time complexity for CIS to roughly $\mathcal{O}(N^3)$ or lower. This polynomial scaling was observed in Fig.~\ref{fig:timings}A. 
We can see that PySEQM’s GPU-enabled semi-empirical excited state calculations are remarkably fast, providing over a order of magnitude speedup in performance over CPU-only calculations.  
For instance, computing 20 excited states for Carbon nanotube-100, the largest molecule in our test set with nearly 1000 atoms, takes about \SI{45}{s}. 
By comparison, performing the same calculation on CPU with PySEQM takes \SI{380}{s}, which is over eight times slower. Performing the same calculation ORCA, which is a quantum chemistry software that does not specialize in semi-empirical excited state calculations, takes 78 minutes. We emphasize, however, that PySEQM is a powerful software even on CPU-only calculations, even though the performance pales before GPU calculations.

Fig.~\ref{fig:timings}B depicts how the total time scales with the number of CIS excited states calculated (upto 100 lowest states) for the three molecular systems. 
The timing data includes the time required to calculate the ground state wave function as well, so the data reflect the total time required for excited state calculation in practical calculations. Testing the excited state method in a variety of molecular systems over a range of excited states helps establish how robust PySEQM is for large scale excited states calculations.
In all the three molecular systems, for molecules with less than 200 atoms (CN 10, PDI 1, PDI 2, PDI 3, Dendrimer 1), the time taken remains relatively constant with respect to the number of excited states calculated (all calculations took under \SI{3}{s}). 
This behavior appears to reflect low GPU utilization. In other words, when the GPU load is below a certain threshold, increasing the number of excited states calculated has little impact on overall runtime. 
For molecules in the intermediate regime of 200 to 500 atoms (CN 20, CN 40, Dendrimer 2, Dendrimer 3, PDI 6, PDI 9), there is roughly a linear increase in computation time with an increasing number of states requested.
The largest molecules studied (with over 500 atoms) are PDI 12, Dendrimer 4, CN 60, CN 80, and CN 100. For these molecules, the time taken per additional state stops being smooth and linearly increasing. 
This irregular behavior in calculation time happens because for these bigger molecules, when the available GPU memory is insufficient to hold the growing Krylov subspace, the subspace is collapsed and the iterative procedure is restarted from the current best estimate of eigenstates. This leads to a greater increase in the number of iterhtions and hence the total time. The convergence pattern in residuals is nonmonotonic with jumps in the total time. 
%In fact for the largest molecule we consider in this study, Carbon Nanotube 100 (with 996 atoms), we are not able to compute all 100 eigenstates at once without exceeding GPU memory. 
If there is a routine necessity of calculating a large number of excited states for large systems we will have to modify our implementation to reduce memory usage. One such strategy is using the Lanczos method\cite{cullum_computing_1981}, which is slower than the Davidson algorithm but has the advantage of being low-memory. 
We could also use Wilkinson's shift method \cite{wilkinson_algebraic_1988} to calculate eigenstates in batches by calculating a smaller subset of the required eigenstates, and then shifting the already calculated eigenstates from the search subspace for new eigenstates.
Advances in GPU hardware with increased memory will also address this constraint.

As previously mentioned, the additional advantage of the current implementation using PyTorch in PySEQM is the ability to run calculations in batch mode. Here, the excited states of multiple molecules can be calculated simultaneously in a single run. 
To assess the efficiency and utility of this functionality, we calculated 15 excited states of uracil neuclobase (with 12 atoms),  PDI-1 (with 43 atoms) and CN-10 (with 132 atoms) across various geometries. We performed a BOMD simulation for each of the molecules in the NVE ensemble with a \SI{0.4}{\femto s} time step, then collected the structures from the first $n$ steps (where $n$ = 1,2,4,8,16,32) to form batched inputs of size $n$.
 These batch sizes represent the minimal practical sizes in machine learning applications. 
For each batch size, we calculated 15 CIS excited states. PySEQM already contains an efficient implementation for the calculation of ground state wave function and running BOMD in batch mode, as discussed in Ref.~\citenum{pyseqm:2020:jctc}.
Fig.~\ref{fig:timings}C shows the total time taken for CIS excited state calculations with increasing batch size, running on a GPU. For the smallest molecule, uracil,  the per-molecule speed-up is about $5\times$ for batch sizes of 16 and 32. For the same batch sizes, the intermediate molecule PDI-1 showed a speed-up of about $3.5\times$, while the biggest molecule, CN-10, had a speed-up of about $2\times$.
Although batch mode can substantially improve GPU utilization, especially for very small systems and when backpropogation is used to calculate forces\cite{pyseqm:2020:jctc}, the gains as seen in Fig.~\ref{fig:timings}C are modest as the system size increases (with smaller speed-ups anticipated for even bigger systems).
Speed-ups achieved from batching will decline with system size as the GPU is generally better utilized, and parallelizing the implementation of the SEQM method becomes more challenging. 

Nonetheless, batching capability still provides a more streamlined workflow that will be useful in machine learning applications and for running ensemble-based nonadiabatic dynamics methods (e.g., surface hopping). 
Even moderate speed-ups can reduce overall costs in large-scale simulations, making batch-mode calculations ideal for nonadiabatic molecular dynamics to estimate reaction rates, lifetimes, product ratios, quantum yields, and other dynamic properties. 

PySEQM is an open-source software made available on Github\cite{noauthor_lanlpyseqm_nodate}.

\section{Conclusion}
We have introduced a PyTorch-based implementation of excited-state calculations for NDDO-based SEQM methods, extending the capabilities of PySEQM software. By leveraging GPU acceleration, our implementation enables excited-state calculation of twenty states at the semi-empirical level for molecular systems with around a thousand atoms in under a minute.

Additionally, batch mode execution enables parallel simulations of multiple molecules, maximizing GPU parallelism. This feature is particularly beneficial for nonadiabatic dynamics simulations, which require running multiple trajectories in parallel.

The PyTorch-based framework also directly supports machine learning applications. The parameters used in SEQM can be re-optimized using PyTorch optimization engines or dynamically corrected via neural networks\cite{zhou_deep_2022}, and our implementation extends this capability to excited states.
Additionally, PySEQM can interface with neural network architectures such as HIPNN\cite{lubbers_hierarchical_2018} enabling machine learning of excited state properties.

In an upcoming publication, we will detail our ongoing efforts to incorporate excited-state dynamics into PySEQM, paving the way for studying reaction rates in large systems where nonadiabatic effects play a crucial role.

\begin{acknowledgement}
This work is supported by the U.S. Department of Energy, Office of Basic Energy Sciences (FWP LANLE8AN) and by the U.S. Department of Energy through the Los Alamos National Laboratory (LANL). LANL is operated by Triad National Security, LLC, for the National Nuclear Security Administration of the U.S. Department of Energy Contract No. 892333218NCA000001.The work at Los Alamos National Laboratory (LANL) was performed in part at the Center for Integrated Nanotechnologies (CINT).
\end{acknowledgement}

\begin{suppinfo}

This will usually read something like: ``Experimental procedures and
characterization data for all new compounds. The class will
automatically add a sentence pointing to the information on-line:

\end{suppinfo}

% \bibliography{achemso-demo}
\bibliography{references.bib}

\providecommand{\latin}[1]{#1}
\makeatletter
\providecommand{\doi}
  {\begingroup\let\do\@makeother\dospecials
  \catcode`\{=1 \catcode`\}=2 \doi@aux}
\providecommand{\doi@aux}[1]{\endgroup\texttt{#1}}
\makeatother
\providecommand*\mcitethebibliography{\thebibliography}
\csname @ifundefined\endcsname{endmcitethebibliography}  {\let\endmcitethebibliography\endthebibliography}{}
\begin{mcitethebibliography}{94}
\providecommand*\natexlab[1]{#1}
\providecommand*\mciteSetBstSublistMode[1]{}
\providecommand*\mciteSetBstMaxWidthForm[2]{}
\providecommand*\mciteBstWouldAddEndPuncttrue
  {\def\EndOfBibitem{\unskip.}}
\providecommand*\mciteBstWouldAddEndPunctfalse
  {\let\EndOfBibitem\relax}
\providecommand*\mciteSetBstMidEndSepPunct[3]{}
\providecommand*\mciteSetBstSublistLabelBeginEnd[3]{}
\providecommand*\EndOfBibitem{}
\mciteSetBstSublistMode{f}
\mciteSetBstMaxWidthForm{subitem}{(\alph{mcitesubitemcount})}
\mciteSetBstSublistLabelBeginEnd
  {\mcitemaxwidthsubitemform\space}
  {\relax}
  {\relax}

\bibitem[Comeau and Bartlett(1993)Comeau, and Bartlett]{bartlett:1993:eomccsd}
Comeau,~D.~C.; Bartlett,~R.~J. The equation-of-motion coupled-cluster method. {Applications} to open- and closed-shell reference states. \emph{Chemical Physics Letters} \textbf{1993}, \emph{207}, 414--423\relax
\mciteBstWouldAddEndPuncttrue
\mciteSetBstMidEndSepPunct{\mcitedefaultmidpunct}
{\mcitedefaultendpunct}{\mcitedefaultseppunct}\relax
\EndOfBibitem
\bibitem[Kozma \latin{et~al.}()Kozma, Tajti, Demoulin, Izsák, Nooijen, and Szalay]{szalay:2020:NewBenchmarkSet:J.Chem.TheoryComput.}
Kozma,~B.; Tajti,~A.; Demoulin,~B.; Izsák,~R.; Nooijen,~M.; Szalay,~P.~G. A {{New Benchmark Set}} for {{Excitation Energy}} of {{Charge Transfer States}}: {{Systematic Investigation}} of {{Coupled Cluster Type Methods}}. \emph{16}, 4213--4225\relax
\mciteBstWouldAddEndPuncttrue
\mciteSetBstMidEndSepPunct{\mcitedefaultmidpunct}
{\mcitedefaultendpunct}{\mcitedefaultseppunct}\relax
\EndOfBibitem
\bibitem[Schirmer(1982)]{schirmer:1982:ADC:Phys.Rev.A}
Schirmer,~J. Beyond the Random-Phase Approximation: {{A}} New Approximation Scheme for the Polarization Propagator. \emph{Physical Review A} \textbf{1982}, \emph{26}, 2395--2416\relax
\mciteBstWouldAddEndPuncttrue
\mciteSetBstMidEndSepPunct{\mcitedefaultmidpunct}
{\mcitedefaultendpunct}{\mcitedefaultseppunct}\relax
\EndOfBibitem
\bibitem[Trofimov and Schirmer(1995)Trofimov, and Schirmer]{schirmer:1995:ADC2:J.Phys.B}
Trofimov,~A.~B.; Schirmer,~J. An Efficient Polarization Propagator Approach to Valence Electron Excitation Spectra. \emph{Journal of Physics B: Atomic, Molecular and Optical Physics} \textbf{1995}, \emph{28}, 2299\relax
\mciteBstWouldAddEndPuncttrue
\mciteSetBstMidEndSepPunct{\mcitedefaultmidpunct}
{\mcitedefaultendpunct}{\mcitedefaultseppunct}\relax
\EndOfBibitem
\bibitem[Dreuw and Wormit(2015)Dreuw, and Wormit]{wormit:2015:ADC:WIREsComput.Mol.Sci.}
Dreuw,~A.; Wormit,~M. The Algebraic Diagrammatic Construction Scheme for the Polarization Propagator for the Calculation of Excited States. \emph{WIREs Computational Molecular Science} \textbf{2015}, \emph{5}, 82--95\relax
\mciteBstWouldAddEndPuncttrue
\mciteSetBstMidEndSepPunct{\mcitedefaultmidpunct}
{\mcitedefaultendpunct}{\mcitedefaultseppunct}\relax
\EndOfBibitem
\bibitem[Roos \latin{et~al.}(1980)Roos, Taylor, and Sigbahn]{sigbahn:1980:casscf:Chem.Phys.}
Roos,~B.~O.; Taylor,~P.~R.; Sigbahn,~P. E.~M. A Complete Active Space {{SCF}} Method ({{CASSCF}}) Using a Density Matrix Formulated Super-{{CI}} Approach. \emph{Chemical Physics} \textbf{1980}, \emph{48}, 157--173\relax
\mciteBstWouldAddEndPuncttrue
\mciteSetBstMidEndSepPunct{\mcitedefaultmidpunct}
{\mcitedefaultendpunct}{\mcitedefaultseppunct}\relax
\EndOfBibitem
\bibitem[Andersson \latin{et~al.}(1990)Andersson, Malmqvist, Roos, Sadlej, and Wolinski]{wolinski:1990:caspt2:J.Phys.Chem.}
Andersson,~{\relax Kerstin}.; Malmqvist,~P.~A.; Roos,~B.~O.; Sadlej,~A.~J.; Wolinski,~{\relax Krzysztof}. Second-Order Perturbation Theory with a {{CASSCF}} Reference Function. \emph{The Journal of Physical Chemistry} \textbf{1990}, \emph{94}, 5483--5488\relax
\mciteBstWouldAddEndPuncttrue
\mciteSetBstMidEndSepPunct{\mcitedefaultmidpunct}
{\mcitedefaultendpunct}{\mcitedefaultseppunct}\relax
\EndOfBibitem
\bibitem[Andersson \latin{et~al.}(1992)Andersson, Malmqvist, and Roos]{roos:1992:caspt2_full:J.Chem.Phys.}
Andersson,~K.; Malmqvist,~P.-{\AA}.; Roos,~B.~O. Second-order Perturbation Theory with a Complete Active Space Self-consistent Field Reference Function. \emph{The Journal of Chemical Physics} \textbf{1992}, \emph{96}, 1218--1226\relax
\mciteBstWouldAddEndPuncttrue
\mciteSetBstMidEndSepPunct{\mcitedefaultmidpunct}
{\mcitedefaultendpunct}{\mcitedefaultseppunct}\relax
\EndOfBibitem
\bibitem[Runge and Gross(1984)Runge, and Gross]{gross:1984:tddft_prl}
Runge,~E.; Gross,~E. K.~U. Density-Functional Theory for Time-Dependent Systems. \emph{Phys. Rev. Lett.} \textbf{1984}, \emph{52}, 997--1000\relax
\mciteBstWouldAddEndPuncttrue
\mciteSetBstMidEndSepPunct{\mcitedefaultmidpunct}
{\mcitedefaultendpunct}{\mcitedefaultseppunct}\relax
\EndOfBibitem
\bibitem[Furche(2001)]{furche:2001:tddft_jcp}
Furche,~F. On the density matrix based approach to time-dependent density functional response theory. \emph{The Journal of Chemical Physics} \textbf{2001}, \emph{114}, 5982--5992\relax
\mciteBstWouldAddEndPuncttrue
\mciteSetBstMidEndSepPunct{\mcitedefaultmidpunct}
{\mcitedefaultendpunct}{\mcitedefaultseppunct}\relax
\EndOfBibitem
\bibitem[Casida()]{casida:tddft:book}
Casida,~M.~E. \emph{Recent Advances in Density Functional Methods}; pp 155--192\relax
\mciteBstWouldAddEndPuncttrue
\mciteSetBstMidEndSepPunct{\mcitedefaultmidpunct}
{\mcitedefaultendpunct}{\mcitedefaultseppunct}\relax
\EndOfBibitem
\bibitem[Shao \latin{et~al.}(2020)Shao, Mei, Sundholm, and Kaila]{yihan:2020:tddft_benchmark}
Shao,~Y.; Mei,~Y.; Sundholm,~D.; Kaila,~V. R.~I. Benchmarking the Performance of Time-Dependent Density Functional Theory Methods on Biochromophores. \emph{Journal of Chemical Theory and Computation} \textbf{2020}, \emph{16}, 587--600\relax
\mciteBstWouldAddEndPuncttrue
\mciteSetBstMidEndSepPunct{\mcitedefaultmidpunct}
{\mcitedefaultendpunct}{\mcitedefaultseppunct}\relax
\EndOfBibitem
\bibitem[Jacquemin \latin{et~al.}(2009)Jacquemin, Wathelet, Perpète, and Adamo]{jctc:2009:tddft_benchmark}
Jacquemin,~D.; Wathelet,~V.; Perpète,~E.~A.; Adamo,~C. Extensive TD-DFT Benchmark: Singlet-Excited States of Organic Molecules. \emph{Journal of Chemical Theory and Computation} \textbf{2009}, \emph{5}, 2420--2435\relax
\mciteBstWouldAddEndPuncttrue
\mciteSetBstMidEndSepPunct{\mcitedefaultmidpunct}
{\mcitedefaultendpunct}{\mcitedefaultseppunct}\relax
\EndOfBibitem
\bibitem[Isegawa \latin{et~al.}(2012)Isegawa, Peverati, and Truhlar]{truhlar:2012:tddft_benchmark}
Isegawa,~M.; Peverati,~R.; Truhlar,~D.~G. Performance of recent and high-performance approximate density functionals for time-dependent density functional theory calculations of valence and Rydberg electronic transition energies. \emph{The Journal of Chemical Physics} \textbf{2012}, \emph{137}, 244104\relax
\mciteBstWouldAddEndPuncttrue
\mciteSetBstMidEndSepPunct{\mcitedefaultmidpunct}
{\mcitedefaultendpunct}{\mcitedefaultseppunct}\relax
\EndOfBibitem
\bibitem[Send \latin{et~al.}(2011)Send, K{\"u}hn, and Furche]{furche:2011:tddft_benchmak:jctc}
Send,~R.; K{\"u}hn,~M.; Furche,~F. Assessing Excited State Methods by Adiabatic Excitation Energies. \emph{Journal of Chemical Theory and Computation} \textbf{2011}, \emph{7}, 2376--2386\relax
\mciteBstWouldAddEndPuncttrue
\mciteSetBstMidEndSepPunct{\mcitedefaultmidpunct}
{\mcitedefaultendpunct}{\mcitedefaultseppunct}\relax
\EndOfBibitem
\bibitem[Mardirossian \latin{et~al.}(2011)Mardirossian, Parkhill, and Head-Gordon]{headgordon:2011:tddft_benchmark}
Mardirossian,~N.; Parkhill,~J.~A.; Head-Gordon,~M. Benchmark results for empirical post-GGA functionals: Difficult exchange problems and independent tests. \emph{Phys. Chem. Chem. Phys.} \textbf{2011}, \emph{13}, 19325--19337\relax
\mciteBstWouldAddEndPuncttrue
\mciteSetBstMidEndSepPunct{\mcitedefaultmidpunct}
{\mcitedefaultendpunct}{\mcitedefaultseppunct}\relax
\EndOfBibitem
\bibitem[Leang \latin{et~al.}(2012)Leang, Zahariev, and Gordon]{mark_gordon:2012:tddft_benchmark}
Leang,~S.~S.; Zahariev,~F.; Gordon,~M.~S. Benchmarking the performance of time-dependent density functional methods. \emph{The Journal of Chemical Physics} \textbf{2012}, \emph{136}, 104101\relax
\mciteBstWouldAddEndPuncttrue
\mciteSetBstMidEndSepPunct{\mcitedefaultmidpunct}
{\mcitedefaultendpunct}{\mcitedefaultseppunct}\relax
\EndOfBibitem
\bibitem[Barca \latin{et~al.}(2018)Barca, Gilbert, and Gill]{barca_simple_2018}
Barca,~G. M.~J.; Gilbert,~A. T.~B.; Gill,~P. M.~W. Simple {Models} for {Difficult} {Electronic} {Excitations}. \emph{Journal of Chemical Theory and Computation} \textbf{2018}, \emph{14}, 1501--1509, Publisher: American Chemical Society\relax
\mciteBstWouldAddEndPuncttrue
\mciteSetBstMidEndSepPunct{\mcitedefaultmidpunct}
{\mcitedefaultendpunct}{\mcitedefaultseppunct}\relax
\EndOfBibitem
\bibitem[Hait and Head-Gordon(2020)Hait, and Head-Gordon]{hait_highly_2020}
Hait,~D.; Head-Gordon,~M. Highly {Accurate} {Prediction} of {Core} {Spectra} of {Molecules} at {Density} {Functional} {Theory} {Cost}: {Attaining} {Sub}-electronvolt {Error} from a {Restricted} {Open}-{Shell} {Kohn}–{Sham} {Approach}. \emph{The Journal of Physical Chemistry Letters} \textbf{2020}, \emph{11}, 775--786, Publisher: American Chemical Society\relax
\mciteBstWouldAddEndPuncttrue
\mciteSetBstMidEndSepPunct{\mcitedefaultmidpunct}
{\mcitedefaultendpunct}{\mcitedefaultseppunct}\relax
\EndOfBibitem
\bibitem[Carter-Fenk and Herbert(2020)Carter-Fenk, and Herbert]{carter-fenk_state-targeted_2020}
Carter-Fenk,~K.; Herbert,~J.~M. State-{Targeted} {Energy} {Projection}: {A} {Simple} and {Robust} {Approach} to {Orbital} {Relaxation} of {Non}-{Aufbau} {Self}-{Consistent} {Field} {Solutions}. \emph{Journal of Chemical Theory and Computation} \textbf{2020}, \emph{16}, 5067--5082, Publisher: American Chemical Society\relax
\mciteBstWouldAddEndPuncttrue
\mciteSetBstMidEndSepPunct{\mcitedefaultmidpunct}
{\mcitedefaultendpunct}{\mcitedefaultseppunct}\relax
\EndOfBibitem
\bibitem[Kunze \latin{et~al.}(2021)Kunze, Hansen, Grimme, and Mewes]{kunze_pcm-roks_2021}
Kunze,~L.; Hansen,~A.; Grimme,~S.; Mewes,~J.-M. {PCM}-{ROKS} for the {Description} of {Charge}-{Transfer} {States} in {Solution}: {Singlet}–{Triplet} {Gaps} with {Chemical} {Accuracy} from {Open}-{Shell} {Kohn}–{Sham} {Reaction}-{Field} {Calculations}. \emph{The Journal of Physical Chemistry Letters} \textbf{2021}, \emph{12}, 8470--8480, Publisher: American Chemical Society\relax
\mciteBstWouldAddEndPuncttrue
\mciteSetBstMidEndSepPunct{\mcitedefaultmidpunct}
{\mcitedefaultendpunct}{\mcitedefaultseppunct}\relax
\EndOfBibitem
\bibitem[Christensen \latin{et~al.}(2016)Christensen, Kubař, Cui, and Elstner]{seqm_review:2016:chem_rev}
Christensen,~A.~S.; Kubař,~T.; Cui,~Q.; Elstner,~M. Semiempirical Quantum Mechanical Methods for Noncovalent Interactions for Chemical and Biochemical Applications. \emph{Chemical Reviews} \textbf{2016}, \emph{116}, 5301--5337\relax
\mciteBstWouldAddEndPuncttrue
\mciteSetBstMidEndSepPunct{\mcitedefaultmidpunct}
{\mcitedefaultendpunct}{\mcitedefaultseppunct}\relax
\EndOfBibitem
\bibitem[Niehaus \latin{et~al.}(2001)Niehaus, Suhai, Della~Sala, Lugli, Elstner, Seifert, and Frauenheim]{niehaus_tight-binding_2001}
Niehaus,~T.~A.; Suhai,~S.; Della~Sala,~F.; Lugli,~P.; Elstner,~M.; Seifert,~G.; Frauenheim,~T. Tight-binding approach to time-dependent density-functional response theory. \emph{Physical Review B} \textbf{2001}, \emph{63}, 085108\relax
\mciteBstWouldAddEndPuncttrue
\mciteSetBstMidEndSepPunct{\mcitedefaultmidpunct}
{\mcitedefaultendpunct}{\mcitedefaultseppunct}\relax
\EndOfBibitem
\bibitem[Domínguez \latin{et~al.}(2013)Domínguez, Aradi, Frauenheim, Lutsker, and Niehaus]{dominguez_extensions_2013}
Domínguez,~A.; Aradi,~B.; Frauenheim,~T.; Lutsker,~V.; Niehaus,~T.~A. Extensions of the {Time}-{Dependent} {Density} {Functional} {Based} {Tight}-{Binding} {Approach}. \emph{Journal of Chemical Theory and Computation} \textbf{2013}, \emph{9}, 4901--4914\relax
\mciteBstWouldAddEndPuncttrue
\mciteSetBstMidEndSepPunct{\mcitedefaultmidpunct}
{\mcitedefaultendpunct}{\mcitedefaultseppunct}\relax
\EndOfBibitem
\bibitem[Nishimoto(2015)]{nishimoto_time-dependent_2015}
Nishimoto,~Y. Time-dependent density-functional tight-binding method with the third-order expansion of electron density. \emph{The Journal of Chemical Physics} \textbf{2015}, \emph{143}, 094108\relax
\mciteBstWouldAddEndPuncttrue
\mciteSetBstMidEndSepPunct{\mcitedefaultmidpunct}
{\mcitedefaultendpunct}{\mcitedefaultseppunct}\relax
\EndOfBibitem
\bibitem[Kranz \latin{et~al.}(2017)Kranz, Elstner, Aradi, Frauenheim, Lutsker, Garcia, and Niehaus]{kranz_time-dependent_2017}
Kranz,~J.~J.; Elstner,~M.; Aradi,~B.; Frauenheim,~T.; Lutsker,~V.; Garcia,~A.~D.; Niehaus,~T.~A. Time-{Dependent} {Extension} of the {Long}-{Range} {Corrected} {Density} {Functional} {Based} {Tight}-{Binding} {Method}. \emph{Journal of Chemical Theory and Computation} \textbf{2017}, \emph{13}, 1737--1747\relax
\mciteBstWouldAddEndPuncttrue
\mciteSetBstMidEndSepPunct{\mcitedefaultmidpunct}
{\mcitedefaultendpunct}{\mcitedefaultseppunct}\relax
\EndOfBibitem
\bibitem[Lutsker \latin{et~al.}(2015)Lutsker, Aradi, and Niehaus]{lutsker_implementation_2015}
Lutsker,~V.; Aradi,~B.; Niehaus,~T.~A. Implementation and benchmark of a long-range corrected functional in the density functional based tight-binding method. \emph{The Journal of Chemical Physics} \textbf{2015}, \emph{143}, 184107\relax
\mciteBstWouldAddEndPuncttrue
\mciteSetBstMidEndSepPunct{\mcitedefaultmidpunct}
{\mcitedefaultendpunct}{\mcitedefaultseppunct}\relax
\EndOfBibitem
\bibitem[Nishimoto(2019)]{nishimoto_time-dependent_2019}
Nishimoto,~Y. Time-{Dependent} {Long}-{Range}-{Corrected} {Density}-{Functional} {Tight}-{Binding} {Method} {Combined} with the {Polarizable} {Continuum} {Model}. \emph{The Journal of Physical Chemistry A} \textbf{2019}, \emph{123}, 5649--5659, Publisher: American Chemical Society\relax
\mciteBstWouldAddEndPuncttrue
\mciteSetBstMidEndSepPunct{\mcitedefaultmidpunct}
{\mcitedefaultendpunct}{\mcitedefaultseppunct}\relax
\EndOfBibitem
\bibitem[Fihey and Jacquemin(2019)Fihey, and Jacquemin]{fihey_performances_2019}
Fihey,~A.; Jacquemin,~D. Performances of {Density} {Functional} {Tight}-{Binding} {Methods} for {Describing} {Ground} and {Excited} {State} {Geometries} of {Organic} {Molecules}. \emph{Journal of Chemical Theory and Computation} \textbf{2019}, \emph{15}, 6267--6276\relax
\mciteBstWouldAddEndPuncttrue
\mciteSetBstMidEndSepPunct{\mcitedefaultmidpunct}
{\mcitedefaultendpunct}{\mcitedefaultseppunct}\relax
\EndOfBibitem
\bibitem[Bertoni and Sánchez(2023)Bertoni, and Sánchez]{bertoni_data-driven_2023}
Bertoni,~A.~I.; Sánchez,~C.~G. Data-driven approach for benchmarking {DFTB}-approximate excited state methods. \emph{Physical Chemistry Chemical Physics} \textbf{2023}, \emph{25}, 3789--3798\relax
\mciteBstWouldAddEndPuncttrue
\mciteSetBstMidEndSepPunct{\mcitedefaultmidpunct}
{\mcitedefaultendpunct}{\mcitedefaultseppunct}\relax
\EndOfBibitem
\bibitem[Koslowski \latin{et~al.}(2003)Koslowski, Beck, and Thiel]{koslowski_implementation_2003}
Koslowski,~A.; Beck,~M.~E.; Thiel,~W. Implementation of a general multireference configuration interaction procedure with analytic gradients in a semiempirical context using the graphical unitary group approach. \emph{Journal of Computational Chemistry} \textbf{2003}, \emph{24}, 714--726, \_eprint: https://onlinelibrary.wiley.com/doi/pdf/10.1002/jcc.10210\relax
\mciteBstWouldAddEndPuncttrue
\mciteSetBstMidEndSepPunct{\mcitedefaultmidpunct}
{\mcitedefaultendpunct}{\mcitedefaultseppunct}\relax
\EndOfBibitem
\bibitem[Kolb and Thiel(1993)Kolb, and Thiel]{kolb_beyond_1993}
Kolb,~M.; Thiel,~W. Beyond the {MNDO} model: {Methodical} considerations and numerical results. \emph{Journal of Computational Chemistry} \textbf{1993}, \emph{14}, 775--789, \_eprint: https://onlinelibrary.wiley.com/doi/pdf/10.1002/jcc.540140704\relax
\mciteBstWouldAddEndPuncttrue
\mciteSetBstMidEndSepPunct{\mcitedefaultmidpunct}
{\mcitedefaultendpunct}{\mcitedefaultseppunct}\relax
\EndOfBibitem
\bibitem[Weber and Thiel(2000)Weber, and Thiel]{weber_orthogonalization_2000}
Weber,~W.; Thiel,~W. Orthogonalization corrections for semiempirical methods. \emph{Theoretical Chemistry Accounts} \textbf{2000}, \emph{103}, 495--506\relax
\mciteBstWouldAddEndPuncttrue
\mciteSetBstMidEndSepPunct{\mcitedefaultmidpunct}
{\mcitedefaultendpunct}{\mcitedefaultseppunct}\relax
\EndOfBibitem
\bibitem[Silva-Junior and Thiel(2010)Silva-Junior, and Thiel]{silva-junior_benchmark_2010}
Silva-Junior,~M.~R.; Thiel,~W. Benchmark of {Electronically} {Excited} {States} for {Semiempirical} {Methods}: {MNDO}, {AM1}, {PM3}, {OM1}, {OM2}, {OM3}, {INDO}/{S}, and {INDO}/{S2}. \emph{Journal of Chemical Theory and Computation} \textbf{2010}, \emph{6}, 1546--1564, Publisher: American Chemical Society\relax
\mciteBstWouldAddEndPuncttrue
\mciteSetBstMidEndSepPunct{\mcitedefaultmidpunct}
{\mcitedefaultendpunct}{\mcitedefaultseppunct}\relax
\EndOfBibitem
\bibitem[Dral \latin{et~al.}(2016)Dral, Wu, Spörkel, Koslowski, Weber, Steiger, Scholten, and Thiel]{dral_semiempirical_2016}
Dral,~P.~O.; Wu,~X.; Spörkel,~L.; Koslowski,~A.; Weber,~W.; Steiger,~R.; Scholten,~M.; Thiel,~W. Semiempirical {Quantum}-{Chemical} {Orthogonalization}-{Corrected} {Methods}: {Theory}, {Implementation}, and {Parameters}. \emph{Journal of Chemical Theory and Computation} \textbf{2016}, \emph{12}, 1082--1096, Publisher: American Chemical Society\relax
\mciteBstWouldAddEndPuncttrue
\mciteSetBstMidEndSepPunct{\mcitedefaultmidpunct}
{\mcitedefaultendpunct}{\mcitedefaultseppunct}\relax
\EndOfBibitem
\bibitem[Liu and Thiel(2018)Liu, and Thiel]{liu_efficient_2018}
Liu,~J.; Thiel,~W. An efficient implementation of semiempirical quantum-chemical orthogonalization-corrected methods for excited-state dynamics. \emph{The Journal of Chemical Physics} \textbf{2018}, \emph{148}, 154103\relax
\mciteBstWouldAddEndPuncttrue
\mciteSetBstMidEndSepPunct{\mcitedefaultmidpunct}
{\mcitedefaultendpunct}{\mcitedefaultseppunct}\relax
\EndOfBibitem
\bibitem[Zobel and González(2021)Zobel, and González]{zobel_quest_2021}
Zobel,~J.~P.; González,~L. The {Quest} to {Simulate} {Excited}-{State} {Dynamics} of {Transition} {Metal} {Complexes}. \emph{JACS Au} \textbf{2021}, \emph{1}, 1116--1140, Publisher: American Chemical Society\relax
\mciteBstWouldAddEndPuncttrue
\mciteSetBstMidEndSepPunct{\mcitedefaultmidpunct}
{\mcitedefaultendpunct}{\mcitedefaultseppunct}\relax
\EndOfBibitem
\bibitem[Westermayr \latin{et~al.}(2019)Westermayr, Gastegger, Menger, Mai, González, and Marquetand]{westermayr_machine_2019}
Westermayr,~J.; Gastegger,~M.; Menger,~M. F. S.~J.; Mai,~S.; González,~L.; Marquetand,~P. Machine learning enables long time scale molecular photodynamics simulations. \emph{Chemical Science} \textbf{2019}, \emph{10}, 8100--8107, Publisher: The Royal Society of Chemistry\relax
\mciteBstWouldAddEndPuncttrue
\mciteSetBstMidEndSepPunct{\mcitedefaultmidpunct}
{\mcitedefaultendpunct}{\mcitedefaultseppunct}\relax
\EndOfBibitem
\bibitem[Akimov(2021)]{akimov_excited_2021}
Akimov,~A.~V. Excited state dynamics in monolayer black phosphorus revisited: {Accounting} for many-body effects. \emph{The Journal of Chemical Physics} \textbf{2021}, \emph{155}, 134106\relax
\mciteBstWouldAddEndPuncttrue
\mciteSetBstMidEndSepPunct{\mcitedefaultmidpunct}
{\mcitedefaultendpunct}{\mcitedefaultseppunct}\relax
\EndOfBibitem
\bibitem[Patel and Bittner(2020)Patel, and Bittner]{patel_mixed_2020}
Patel,~K.; Bittner,~E.~R. Mixed {Quantum} {Classical} {Simulations} of {Charge}-{Transfer} {Dynamics} in a {Model} {Light}-{Harvesting} {Complex}. {I}. {Charge}-{Transfer} {Dynamics}. \emph{The Journal of Physical Chemistry B} \textbf{2020}, \emph{124}, 2149--2157, Publisher: American Chemical Society\relax
\mciteBstWouldAddEndPuncttrue
\mciteSetBstMidEndSepPunct{\mcitedefaultmidpunct}
{\mcitedefaultendpunct}{\mcitedefaultseppunct}\relax
\EndOfBibitem
\bibitem[Pal \latin{et~al.}(2016)Pal, Trivedi, Akimov, Aradi, Frauenheim, and Prezhdo]{pal_nonadiabatic_2016}
Pal,~S.; Trivedi,~D.~J.; Akimov,~A.~V.; Aradi,~B.; Frauenheim,~T.; Prezhdo,~O.~V. Nonadiabatic {Molecular} {Dynamics} for {Thousand} {Atom} {Systems}: {A} {Tight}-{Binding} {Approach} toward {PYXAID}. \emph{Journal of Chemical Theory and Computation} \textbf{2016}, \emph{12}, 1436--1448, Publisher: American Chemical Society\relax
\mciteBstWouldAddEndPuncttrue
\mciteSetBstMidEndSepPunct{\mcitedefaultmidpunct}
{\mcitedefaultendpunct}{\mcitedefaultseppunct}\relax
\EndOfBibitem
\bibitem[Ufimtsev and Mart{\'\i}nez(2008)Ufimtsev, and Mart{\'\i}nez]{TMartinez08}
Ufimtsev,~I.~S.; Mart{\'\i}nez,~T.~J. {Quantum chemistry on graphical processing units. 1. Strategies for two-electron integral evaluation}. \emph{J. Chem. Theory Comput.} \textbf{2008}, \emph{4}, 222--231\relax
\mciteBstWouldAddEndPuncttrue
\mciteSetBstMidEndSepPunct{\mcitedefaultmidpunct}
{\mcitedefaultendpunct}{\mcitedefaultseppunct}\relax
\EndOfBibitem
\bibitem[Germann \latin{et~al.}(2009)Germann, Kadau, and Swaminarayan]{TGermann09}
Germann,~T.~C.; Kadau,~K.; Swaminarayan,~S. 369 {Tflop}/s molecular dynamics simulations on the petaflop hybrid supercomputer `{Roadrunner}'. \emph{Concurrency and Computation: Practice and Experience} \textbf{2009}, \emph{21}, 2143--2159\relax
\mciteBstWouldAddEndPuncttrue
\mciteSetBstMidEndSepPunct{\mcitedefaultmidpunct}
{\mcitedefaultendpunct}{\mcitedefaultseppunct}\relax
\EndOfBibitem
\bibitem[Ufimtsev and Mart{\'\i}nez(2009)Ufimtsev, and Mart{\'\i}nez]{TMartinez09a}
Ufimtsev,~I.~S.; Mart{\'\i}nez,~T.~J. Quantum Chemistry on Graphical Processing Units. 2. Direct Self-Consistent-Field Implementation. \emph{J. Chem. Theory Comput.} \textbf{2009}, \emph{5}, 1004--1015\relax
\mciteBstWouldAddEndPuncttrue
\mciteSetBstMidEndSepPunct{\mcitedefaultmidpunct}
{\mcitedefaultendpunct}{\mcitedefaultseppunct}\relax
\EndOfBibitem
\bibitem[Ufimtsev and Mart{\'\i}nez(2009)Ufimtsev, and Mart{\'\i}nez]{TMartinez09b}
Ufimtsev,~I.~S.; Mart{\'\i}nez,~T.~J. {Quantum chemistry on graphical processing units. 3. Analytical energy gradients, geometry optimization, and first principles molecular dynamics}. \emph{J. Chem. Theory Comput.} \textbf{2009}, \emph{5}, 2619--2628\relax
\mciteBstWouldAddEndPuncttrue
\mciteSetBstMidEndSepPunct{\mcitedefaultmidpunct}
{\mcitedefaultendpunct}{\mcitedefaultseppunct}\relax
\EndOfBibitem
\bibitem[Genovese \latin{et~al.}(2009)Genovese, Ospici, Deutsch, M{\'e}haut, Neelov, and Goedecker]{SGoedecker09}
Genovese,~L.; Ospici,~M.; Deutsch,~T.; M{\'e}haut,~J.-F.; Neelov,~A.; Goedecker,~S. Density functional theory calculation on many-cores hybrid central processing unit-graphic processing unit architectures. \emph{J. Chem. Phys.} \textbf{2009}, \emph{131}, 034103\relax
\mciteBstWouldAddEndPuncttrue
\mciteSetBstMidEndSepPunct{\mcitedefaultmidpunct}
{\mcitedefaultendpunct}{\mcitedefaultseppunct}\relax
\EndOfBibitem
\bibitem[Stone \latin{et~al.}(2010)Stone, Hardy, Ufimtsev, and Schulten]{JStone10}
Stone,~J.~E.; Hardy,~D.~J.; Ufimtsev,~I.~S.; Schulten,~K. {GPU-accelerated molecular modeling coming of age}. \emph{J. Mol. Graph. Model.} \textbf{2010}, \emph{29}, 116 -- 125\relax
\mciteBstWouldAddEndPuncttrue
\mciteSetBstMidEndSepPunct{\mcitedefaultmidpunct}
{\mcitedefaultendpunct}{\mcitedefaultseppunct}\relax
\EndOfBibitem
\bibitem[Luehr \latin{et~al.}(2011)Luehr, Ufimtsev, and Mart{\'\i}nez]{TMartinez11}
Luehr,~N.; Ufimtsev,~I.~S.; Mart{\'\i}nez,~T.~J. Dynamic Precision for Electron Repulsion Integral Evaluation on Graphical Processing Units ({GPU}s). \emph{J. Chem. Theory Comput.} \textbf{2011}, \emph{7}, 949--954\relax
\mciteBstWouldAddEndPuncttrue
\mciteSetBstMidEndSepPunct{\mcitedefaultmidpunct}
{\mcitedefaultendpunct}{\mcitedefaultseppunct}\relax
\EndOfBibitem
\bibitem[Maia \latin{et~al.}(2015)Maia, Carvalho, Mangueira, Santana, Cabral, and Rocha]{JMaia12}
Maia,~J. D.~C.; Carvalho,~G. A.~U.; Mangueira,~C.~P.; Santana,~S.~R.; Cabral,~L. A.~F.; Rocha,~G.~B. {GPU} Linear Algebra Libraries and {GPGPU} Programming for Accelerating {MOPAC} Semiempirical Quantum Chemistry Calculations. \emph{J. Chem. Theory Comput.} \textbf{2015}, \emph{11}, 3131--3144\relax
\mciteBstWouldAddEndPuncttrue
\mciteSetBstMidEndSepPunct{\mcitedefaultmidpunct}
{\mcitedefaultendpunct}{\mcitedefaultseppunct}\relax
\EndOfBibitem
\bibitem[Hacene \latin{et~al.}(2012)Hacene, Anciaux-Sedrakian, Rozanska, Klahr, Guignon, and Fleurat-Lessard]{MHacene12}
Hacene,~M.; Anciaux-Sedrakian,~A.; Rozanska,~X.; Klahr,~D.; Guignon,~T.; Fleurat-Lessard,~P. {Accelerating VASP electronic structure calculations using graphic processing units}. \emph{J. Chem. Theory Comput.} \textbf{2012}, \emph{33}, 2581--2589\relax
\mciteBstWouldAddEndPuncttrue
\mciteSetBstMidEndSepPunct{\mcitedefaultmidpunct}
{\mcitedefaultendpunct}{\mcitedefaultseppunct}\relax
\EndOfBibitem
\bibitem[Cawkwell \latin{et~al.}(2012)Cawkwell, Sanville, Mniszewski, and Niklasson]{MCawkwell12b}
Cawkwell,~M.~J.; Sanville,~E.~J.; Mniszewski,~S.~M.; Niklasson,~A. M.~N. Computing the Density Matrix in Electronic Structure Theory on Graphics Processing Units. \emph{Journal of Chemical Theory and Computation} \textbf{2012}, \emph{8}, 4094--4101\relax
\mciteBstWouldAddEndPuncttrue
\mciteSetBstMidEndSepPunct{\mcitedefaultmidpunct}
{\mcitedefaultendpunct}{\mcitedefaultseppunct}\relax
\EndOfBibitem
\bibitem[Liu \latin{et~al.}(2012)Liu, Luehr, Kulik, and Mart{\'\i}nez]{FLiu2015}
Liu,~F.; Luehr,~N.; Kulik,~H.~J.; Mart{\'\i}nez,~T.~J. Quantum chemistry for solvated molecules on graphical processing units using polarizable continuum models. \emph{Comput. Phys. Commun.} \textbf{2012}, \emph{8}, 3072--3081\relax
\mciteBstWouldAddEndPuncttrue
\mciteSetBstMidEndSepPunct{\mcitedefaultmidpunct}
{\mcitedefaultendpunct}{\mcitedefaultseppunct}\relax
\EndOfBibitem
\bibitem[Huhn \latin{et~al.}(2020)Huhn, Lange, Yu, Yoon, and Blum]{WHuhn20}
Huhn,~W.~P.; Lange,~B.; Yu,~V.~W.; Yoon,~M.; Blum,~V. {GPU} acceleration of all-electron electronic structure theory using localized numeric atom-centered basis functions. \emph{Comput. Phys. Commun.} \textbf{2020}, \emph{254}, 107314\relax
\mciteBstWouldAddEndPuncttrue
\mciteSetBstMidEndSepPunct{\mcitedefaultmidpunct}
{\mcitedefaultendpunct}{\mcitedefaultseppunct}\relax
\EndOfBibitem
\bibitem[Gordon and Windus(2020)Gordon, and Windus]{MGordon20}
Gordon,~M.~S.; Windus,~T.~L. Modern Architectures and Their Impact on Electronic Structure Theory. \emph{Chem. Rev.} \textbf{2020}, \emph{120}, 9015--9020\relax
\mciteBstWouldAddEndPuncttrue
\mciteSetBstMidEndSepPunct{\mcitedefaultmidpunct}
{\mcitedefaultendpunct}{\mcitedefaultseppunct}\relax
\EndOfBibitem
\bibitem[Fales and Levine(2015)Fales, and Levine]{fales_nanoscale_2015}
Fales,~B.~S.; Levine,~B.~G. Nanoscale {Multireference} {Quantum} {Chemistry}: {Full} {Configuration} {Interaction} on {Graphical} {Processing} {Units}. \emph{Journal of Chemical Theory and Computation} \textbf{2015}, \emph{11}, 4708--4716, Publisher: American Chemical Society\relax
\mciteBstWouldAddEndPuncttrue
\mciteSetBstMidEndSepPunct{\mcitedefaultmidpunct}
{\mcitedefaultendpunct}{\mcitedefaultseppunct}\relax
\EndOfBibitem
\bibitem[Zhou \latin{et~al.}(2020)Zhou, Nebgen, Lubbers, Malone, Niklasson, and Tretiak]{ZGuoqing20}
Zhou,~G.; Nebgen,~B.; Lubbers,~N.; Malone,~W.; Niklasson,~A. M.~N.; Tretiak,~S. Graphics Processing Unit-Accelerated Semiempirical Born Oppenheimer Molecular Dynamics Using {PyTorch}. \emph{J. Chem. Theory Comput.} \textbf{2020}, \emph{16}, 4951--4962\relax
\mciteBstWouldAddEndPuncttrue
\mciteSetBstMidEndSepPunct{\mcitedefaultmidpunct}
{\mcitedefaultendpunct}{\mcitedefaultseppunct}\relax
\EndOfBibitem
\bibitem[Finkelstein \latin{et~al.}(2021)Finkelstein, Smith, Mniszewski, Barros, Negre, Rubensson, and Niklasson]{JFinkelstein21}
Finkelstein,~J.; Smith,~J.~S.; Mniszewski,~S.~M.; Barros,~K.; Negre,~C. F.~A.; Rubensson,~E.~H.; Niklasson,~A. M.~N. Mixed Precision Fermi-Operator Expansion on Tensor Cores from a Machine Learning Perspective. \emph{Journal of Chemical Theory and Computation} \textbf{2021}, \emph{17}, 2256--2265\relax
\mciteBstWouldAddEndPuncttrue
\mciteSetBstMidEndSepPunct{\mcitedefaultmidpunct}
{\mcitedefaultendpunct}{\mcitedefaultseppunct}\relax
\EndOfBibitem
\bibitem[Finkelstein \latin{et~al.}(2021)Finkelstein, Smith, Mniszewski, Barros, Negre, Rubensson, and Niklasson]{JFinkelstein21B}
Finkelstein,~J.; Smith,~J.~S.; Mniszewski,~S.~M.; Barros,~K.; Negre,~C. F.~A.; Rubensson,~E.~H.; Niklasson,~A. M.~N. Quantum-Based Molecular Dynamics Simulations Using Tensor Cores. \emph{Journal of Chemical Theory and Computation} \textbf{2021}, \emph{17}, 6180--6192\relax
\mciteBstWouldAddEndPuncttrue
\mciteSetBstMidEndSepPunct{\mcitedefaultmidpunct}
{\mcitedefaultendpunct}{\mcitedefaultseppunct}\relax
\EndOfBibitem
\bibitem[Kim \latin{et~al.}(2023)Kim, Jeong, Son, Kim, Rhee, Jung, Choi, Yim, Jang, and Kim]{kim_kohnsham_2023}
Kim,~I.; Jeong,~D.; Son,~W.-J.; Kim,~H.-J.; Rhee,~Y.~M.; Jung,~Y.; Choi,~H.; Yim,~J.; Jang,~I.; Kim,~D.~S. Kohn–{Sham} time-dependent density functional theory with {Tamm}–{Dancoff} approximation on massively parallel {GPUs}. \emph{npj Computational Materials} \textbf{2023}, \emph{9}, 1--12, Publisher: Nature Publishing Group\relax
\mciteBstWouldAddEndPuncttrue
\mciteSetBstMidEndSepPunct{\mcitedefaultmidpunct}
{\mcitedefaultendpunct}{\mcitedefaultseppunct}\relax
\EndOfBibitem
\bibitem[Kim \latin{et~al.}(2024)Kim, Jeong, Weisburn, Alexiu, Van~Voorhis, Rhee, Son, Kim, Yim, Kim, Cho, Jang, Lee, and Kim]{kim_very-large-scale_2024}
Kim,~I.; Jeong,~D.; Weisburn,~L.~P.; Alexiu,~A.; Van~Voorhis,~T.; Rhee,~Y.~M.; Son,~W.-J.; Kim,~H.-J.; Yim,~J.; Kim,~S.; Cho,~Y.; Jang,~I.; Lee,~S.; Kim,~D.~S. Very-{Large}-{Scale} {GPU}-{Accelerated} {Nuclear} {Gradient} of {Time}-{Dependent} {Density} {Functional} {Theory} with {Tamm}–{Dancoff} {Approximation} and {Range}-{Separated} {Hybrid} {Functionals}. \emph{Journal of Chemical Theory and Computation} \textbf{2024}, \emph{20}, 9018--9031, Publisher: American Chemical Society\relax
\mciteBstWouldAddEndPuncttrue
\mciteSetBstMidEndSepPunct{\mcitedefaultmidpunct}
{\mcitedefaultendpunct}{\mcitedefaultseppunct}\relax
\EndOfBibitem
\bibitem[Zhou \latin{et~al.}(2020)Zhou, Nebgen, Lubbers, Malone, Niklasson, and Tretiak]{pyseqm:2020:jctc}
Zhou,~G.; Nebgen,~B.; Lubbers,~N.; Malone,~W.; Niklasson,~A. M.~N.; Tretiak,~S. Graphics Processing Unit-Accelerated Semiempirical Born Oppenheimer Molecular Dynamics Using PyTorch. \emph{Journal of Chemical Theory and Computation} \textbf{2020}, \emph{16}, 4951--4962\relax
\mciteBstWouldAddEndPuncttrue
\mciteSetBstMidEndSepPunct{\mcitedefaultmidpunct}
{\mcitedefaultendpunct}{\mcitedefaultseppunct}\relax
\EndOfBibitem
\bibitem[Zhou \latin{et~al.}(2022)Zhou, Lubbers, Barros, Tretiak, and Nebgen]{zhou_deep_2022}
Zhou,~G.; Lubbers,~N.; Barros,~K.; Tretiak,~S.; Nebgen,~B. Deep learning of dynamically responsive chemical {Hamiltonians} with semiempirical quantum mechanics. \emph{Proceedings of the National Academy of Sciences} \textbf{2022}, \emph{119}, e2120333119, Publisher: Proceedings of the National Academy of Sciences\relax
\mciteBstWouldAddEndPuncttrue
\mciteSetBstMidEndSepPunct{\mcitedefaultmidpunct}
{\mcitedefaultendpunct}{\mcitedefaultseppunct}\relax
\EndOfBibitem
\bibitem[Dewar and Thiel(1977)Dewar, and Thiel]{dewar_ground_1977}
Dewar,~M. J.~S.; Thiel,~W. Ground states of molecules. 38. {The} {MNDO} method. {Approximations} and parameters. \emph{Journal of the American Chemical Society} \textbf{1977}, \emph{99}, 4899--4907, Publisher: American Chemical Society\relax
\mciteBstWouldAddEndPuncttrue
\mciteSetBstMidEndSepPunct{\mcitedefaultmidpunct}
{\mcitedefaultendpunct}{\mcitedefaultseppunct}\relax
\EndOfBibitem
\bibitem[Dewar \latin{et~al.}(1985)Dewar, Zoebisch, Healy, and Stewart]{dewar_development_1985}
Dewar,~M. J.~S.; Zoebisch,~E.~G.; Healy,~E.~F.; Stewart,~J. J.~P. Development and use of quantum mechanical molecular models. 76. {AM1}: a new general purpose quantum mechanical molecular model. \emph{Journal of the American Chemical Society} \textbf{1985}, \emph{107}, 3902--3909, Publisher: American Chemical Society\relax
\mciteBstWouldAddEndPuncttrue
\mciteSetBstMidEndSepPunct{\mcitedefaultmidpunct}
{\mcitedefaultendpunct}{\mcitedefaultseppunct}\relax
\EndOfBibitem
\bibitem[Stewart(1989)]{stewart_optimization_1989}
Stewart,~J. J.~P. Optimization of parameters for semiempirical methods {I}. {Method}. \emph{Journal of Computational Chemistry} \textbf{1989}, \emph{10}, 209--220, \_eprint: https://onlinelibrary.wiley.com/doi/pdf/10.1002/jcc.540100208\relax
\mciteBstWouldAddEndPuncttrue
\mciteSetBstMidEndSepPunct{\mcitedefaultmidpunct}
{\mcitedefaultendpunct}{\mcitedefaultseppunct}\relax
\EndOfBibitem
\bibitem[Stewart(2007)]{stewart_optimization_2007}
Stewart,~J. J.~P. Optimization of parameters for semiempirical methods {V}: {Modification} of {NDDO} approximations and application to 70 elements. \emph{Journal of Molecular Modeling} \textbf{2007}, \emph{13}, 1173--1213\relax
\mciteBstWouldAddEndPuncttrue
\mciteSetBstMidEndSepPunct{\mcitedefaultmidpunct}
{\mcitedefaultendpunct}{\mcitedefaultseppunct}\relax
\EndOfBibitem
\bibitem[Niklasson(2008)]{niklasson_extended_2008}
Niklasson,~A. M.~N. Extended {Born}-{Oppenheimer} {Molecular} {Dynamics}. \emph{Physical Review Letters} \textbf{2008}, \emph{100}, 123004, Publisher: American Physical Society\relax
\mciteBstWouldAddEndPuncttrue
\mciteSetBstMidEndSepPunct{\mcitedefaultmidpunct}
{\mcitedefaultendpunct}{\mcitedefaultseppunct}\relax
\EndOfBibitem
\bibitem[Niklasson \latin{et~al.}(2009)Niklasson, Steneteg, Odell, Bock, Challacombe, Tymczak, Holmström, Zheng, and Weber]{niklasson_extended_2009}
Niklasson,~A. M.~N.; Steneteg,~P.; Odell,~A.; Bock,~N.; Challacombe,~M.; Tymczak,~C.~J.; Holmström,~E.; Zheng,~G.; Weber,~V. Extended {Lagrangian} {Born}–{Oppenheimer} molecular dynamics with dissipation. \emph{The Journal of Chemical Physics} \textbf{2009}, \emph{130}, 214109\relax
\mciteBstWouldAddEndPuncttrue
\mciteSetBstMidEndSepPunct{\mcitedefaultmidpunct}
{\mcitedefaultendpunct}{\mcitedefaultseppunct}\relax
\EndOfBibitem
\bibitem[Niklasson(2020)]{niklasson_density-matrix_2020}
Niklasson,~A. M.~N. Density-{Matrix} {Based} {Extended} {Lagrangian} {Born}–{Oppenheimer} {Molecular} {Dynamics}. \emph{Journal of Chemical Theory and Computation} \textbf{2020}, \emph{16}, 3628--3640, Publisher: American Chemical Society\relax
\mciteBstWouldAddEndPuncttrue
\mciteSetBstMidEndSepPunct{\mcitedefaultmidpunct}
{\mcitedefaultendpunct}{\mcitedefaultseppunct}\relax
\EndOfBibitem
\bibitem[Niklasson(2021)]{niklasson_extended_2021}
Niklasson,~A. M.~N. Extended {Lagrangian} {Born}–{Oppenheimer} molecular dynamics: from density functional theory to charge relaxation models. \emph{The European Physical Journal B} \textbf{2021}, \emph{94}, 164\relax
\mciteBstWouldAddEndPuncttrue
\mciteSetBstMidEndSepPunct{\mcitedefaultmidpunct}
{\mcitedefaultendpunct}{\mcitedefaultseppunct}\relax
\EndOfBibitem
\bibitem[Kulichenko \latin{et~al.}(2023)Kulichenko, Barros, Lubbers, Fedik, Zhou, Tretiak, Nebgen, and Niklasson]{kulichenko_semi-empirical_2023}
Kulichenko,~M.; Barros,~K.; Lubbers,~N.; Fedik,~N.; Zhou,~G.; Tretiak,~S.; Nebgen,~B.; Niklasson,~A. M.~N. Semi-{Empirical} {Shadow} {Molecular} {Dynamics}: {A} {PyTorch} {Implementation}. \emph{Journal of Chemical Theory and Computation} \textbf{2023}, \emph{19}, 3209--3222, Publisher: American Chemical Society\relax
\mciteBstWouldAddEndPuncttrue
\mciteSetBstMidEndSepPunct{\mcitedefaultmidpunct}
{\mcitedefaultendpunct}{\mcitedefaultseppunct}\relax
\EndOfBibitem
\bibitem[Lubbers \latin{et~al.}(2018)Lubbers, Smith, and Barros]{lubbers_hierarchical_2018}
Lubbers,~N.; Smith,~J.~S.; Barros,~K. Hierarchical modeling of molecular energies using a deep neural network. \emph{The Journal of Chemical Physics} \textbf{2018}, \emph{148}, 241715\relax
\mciteBstWouldAddEndPuncttrue
\mciteSetBstMidEndSepPunct{\mcitedefaultmidpunct}
{\mcitedefaultendpunct}{\mcitedefaultseppunct}\relax
\EndOfBibitem
\bibitem[Dreuw and Head-Gordon(2005)Dreuw, and Head-Gordon]{dreuw_single-reference_2005}
Dreuw,~A.; Head-Gordon,~M. Single-{Reference} ab {Initio} {Methods} for the {Calculation} of {Excited} {States} of {Large} {Molecules}. \emph{Chemical Reviews} \textbf{2005}, \emph{105}, 4009--4037, Publisher: American Chemical Society\relax
\mciteBstWouldAddEndPuncttrue
\mciteSetBstMidEndSepPunct{\mcitedefaultmidpunct}
{\mcitedefaultendpunct}{\mcitedefaultseppunct}\relax
\EndOfBibitem
\bibitem[Davidson(1975)]{davidson_iterative_1975}
Davidson,~E.~R. The iterative calculation of a few of the lowest eigenvalues and corresponding eigenvectors of large real-symmetric matrices. \emph{Journal of Computational Physics} \textbf{1975}, \emph{17}, 87--94\relax
\mciteBstWouldAddEndPuncttrue
\mciteSetBstMidEndSepPunct{\mcitedefaultmidpunct}
{\mcitedefaultendpunct}{\mcitedefaultseppunct}\relax
\EndOfBibitem
\bibitem[Liu(1978)]{liu_simultaneous_1978}
Liu,~B. The simultaneous expansion method for the iterative solution of several of the lowest eigenvalues and corresponding eigenvectors of large real-symmetric matrices. \emph{Numerical algorithms in chemistry: algebraic methods} \textbf{1978}, 49--53, Publisher: Lawrence Berkeley Laboratory, University of California Berkeley\relax
\mciteBstWouldAddEndPuncttrue
\mciteSetBstMidEndSepPunct{\mcitedefaultmidpunct}
{\mcitedefaultendpunct}{\mcitedefaultseppunct}\relax
\EndOfBibitem
\bibitem[Neese(2022)]{neese_software_2022}
Neese,~F. Software update: {The} {ORCA} program system—{Version} 5.0. \emph{WIREs Computational Molecular Science} \textbf{2022}, \emph{12}, e1606, \_eprint: https://onlinelibrary.wiley.com/doi/pdf/10.1002/wcms.1606\relax
\mciteBstWouldAddEndPuncttrue
\mciteSetBstMidEndSepPunct{\mcitedefaultmidpunct}
{\mcitedefaultendpunct}{\mcitedefaultseppunct}\relax
\EndOfBibitem
\bibitem[Peng \latin{et~al.}(2006)Peng, Melinger, and Kleiman]{dendrimer:2006:photoresearch}
Peng,~Z.; Melinger,~J.~S.; Kleiman,~V. Light harvesting unsymmetrical conjugated dendrimders as photosynthetic mimics. \emph{Photosynthesis Research} \textbf{2006}, \emph{87}, 115--131\relax
\mciteBstWouldAddEndPuncttrue
\mciteSetBstMidEndSepPunct{\mcitedefaultmidpunct}
{\mcitedefaultendpunct}{\mcitedefaultseppunct}\relax
\EndOfBibitem
\bibitem[Shortreed \latin{et~al.}(1997)Shortreed, Swallen, Shi, Tan, Xu, Devadoss, Moore, and Kopelman]{shortreed_directed_1997}
Shortreed,~M.~R.; Swallen,~S.~F.; Shi,~Z.-Y.; Tan,~W.; Xu,~Z.; Devadoss,~C.; Moore,~J.~S.; Kopelman,~R. Directed {Energy} {Transfer} {Funnels} in {Dendrimeric} {Antenna} {Supermolecules}. \emph{The Journal of Physical Chemistry B} \textbf{1997}, \emph{101}, 6318--6322, Publisher: American Chemical Society\relax
\mciteBstWouldAddEndPuncttrue
\mciteSetBstMidEndSepPunct{\mcitedefaultmidpunct}
{\mcitedefaultendpunct}{\mcitedefaultseppunct}\relax
\EndOfBibitem
\bibitem[Palma \latin{et~al.}(2010)Palma, Atas, Hardison, Marder, Collings, Beeby, Melinger, Krause, Kleiman, and Roitberg]{palma_electronic_2010}
Palma,~J.~L.; Atas,~E.; Hardison,~L.; Marder,~T.~B.; Collings,~J.~C.; Beeby,~A.; Melinger,~J.~S.; Krause,~J.~L.; Kleiman,~V.~D.; Roitberg,~A.~E. Electronic {Spectra} of the {Nanostar} {Dendrimer}: {Theory} and {Experiment}. \emph{The Journal of Physical Chemistry C} \textbf{2010}, \emph{114}, 20702--20712, Publisher: American Chemical Society\relax
\mciteBstWouldAddEndPuncttrue
\mciteSetBstMidEndSepPunct{\mcitedefaultmidpunct}
{\mcitedefaultendpunct}{\mcitedefaultseppunct}\relax
\EndOfBibitem
\bibitem[Ackermann \latin{et~al.}(2022)Ackermann, Metternich, Herbertz, and Kruss]{ackermann_biosensing_2022}
Ackermann,~J.; Metternich,~J.~T.; Herbertz,~S.; Kruss,~S. Biosensing with {Fluorescent} {Carbon} {Nanotubes}. \emph{Angewandte Chemie International Edition} \textbf{2022}, \emph{61}, e202112372, \_eprint: https://onlinelibrary.wiley.com/doi/pdf/10.1002/anie.202112372\relax
\mciteBstWouldAddEndPuncttrue
\mciteSetBstMidEndSepPunct{\mcitedefaultmidpunct}
{\mcitedefaultendpunct}{\mcitedefaultseppunct}\relax
\EndOfBibitem
\bibitem[Wieland \latin{et~al.}(2021)Wieland, Li, Rust, Chen, and Flavel]{wieland_carbon_2021}
Wieland,~L.; Li,~H.; Rust,~C.; Chen,~J.; Flavel,~B.~S. Carbon {Nanotubes} for {Photovoltaics}: {From} {Lab} to {Industry}. \emph{Advanced Energy Materials} \textbf{2021}, \emph{11}, 2002880, \_eprint: https://onlinelibrary.wiley.com/doi/pdf/10.1002/aenm.202002880\relax
\mciteBstWouldAddEndPuncttrue
\mciteSetBstMidEndSepPunct{\mcitedefaultmidpunct}
{\mcitedefaultendpunct}{\mcitedefaultseppunct}\relax
\EndOfBibitem
\bibitem[Zacheo \latin{et~al.}(2024)Zacheo, Matano, Shimura, Yu, Doumani, Komatsu, Kono, and Maki]{zacheo_efficient_2024}
Zacheo,~A.; Matano,~S.; Shimura,~Y.; Yu,~S.; Doumani,~J.; Komatsu,~N.; Kono,~J.; Maki,~H. Efficient {Emission} of {Highly} {Polarized} {Thermal} {Radiation} from a {Suspended} {Aligned} {Carbon} {Nanotube} {Film}. \emph{ACS Nano} \textbf{2024}, \emph{18}, 15769--15778, Publisher: American Chemical Society\relax
\mciteBstWouldAddEndPuncttrue
\mciteSetBstMidEndSepPunct{\mcitedefaultmidpunct}
{\mcitedefaultendpunct}{\mcitedefaultseppunct}\relax
\EndOfBibitem
\bibitem[Mann \latin{et~al.}(2007)Mann, Kato, Kinkhabwala, Pop, Cao, Wang, Zhang, Wang, Guo, and Dai]{mann_electrically_2007}
Mann,~D.; Kato,~Y.~K.; Kinkhabwala,~A.; Pop,~E.; Cao,~J.; Wang,~X.; Zhang,~L.; Wang,~Q.; Guo,~J.; Dai,~H. Electrically driven thermal light emission from individual single-walled carbon nanotubes. \emph{Nature Nanotechnology} \textbf{2007}, \emph{2}, 33--38, Publisher: Nature Publishing Group\relax
\mciteBstWouldAddEndPuncttrue
\mciteSetBstMidEndSepPunct{\mcitedefaultmidpunct}
{\mcitedefaultendpunct}{\mcitedefaultseppunct}\relax
\EndOfBibitem
\bibitem[Astruc \latin{et~al.}(2010)Astruc, Boisselier, and Ornelas]{astruc_dendrimers_2010}
Astruc,~D.; Boisselier,~E.; Ornelas,~C. Dendrimers {Designed} for {Functions}: {From} {Physical}, {Photophysical}, and {Supramolecular} {Properties} to {Applications} in {Sensing}, {Catalysis}, {Molecular} {Electronics}, {Photonics}, and {Nanomedicine}. \emph{Chemical Reviews} \textbf{2010}, \emph{110}, 1857--1959, Publisher: American Chemical Society\relax
\mciteBstWouldAddEndPuncttrue
\mciteSetBstMidEndSepPunct{\mcitedefaultmidpunct}
{\mcitedefaultendpunct}{\mcitedefaultseppunct}\relax
\EndOfBibitem
\bibitem[Kim and Zimmerman(1998)Kim, and Zimmerman]{kim_applications_1998}
Kim,~Y.; Zimmerman,~S.~C. Applications of dendrimers in bio-organic chemistry. \emph{Current Opinion in Chemical Biology} \textbf{1998}, \emph{2}, 733--742\relax
\mciteBstWouldAddEndPuncttrue
\mciteSetBstMidEndSepPunct{\mcitedefaultmidpunct}
{\mcitedefaultendpunct}{\mcitedefaultseppunct}\relax
\EndOfBibitem
\bibitem[Chaudhuri \latin{et~al.}(2011)Chaudhuri, Li, Che, Shafran, Gerton, Zang, and Lupton]{chaudhuri_enhancing_2011}
Chaudhuri,~D.; Li,~D.; Che,~Y.; Shafran,~E.; Gerton,~J.~M.; Zang,~L.; Lupton,~J.~M. Enhancing {Long}-{Range} {Exciton} {Guiding} in {Molecular} {Nanowires} by {H}-{Aggregation} {Lifetime} {Engineering}. \emph{Nano Letters} \textbf{2011}, \emph{11}, 488--492, Publisher: American Chemical Society\relax
\mciteBstWouldAddEndPuncttrue
\mciteSetBstMidEndSepPunct{\mcitedefaultmidpunct}
{\mcitedefaultendpunct}{\mcitedefaultseppunct}\relax
\EndOfBibitem
\bibitem[Zhou \latin{et~al.}(2018)Zhou, Zhang, Jiang, Wang, Zhou, Xu, Liu, Xie, and Ma]{zhou_magic-angle_2018}
Zhou,~J.; Zhang,~W.; Jiang,~X.-F.; Wang,~C.; Zhou,~X.; Xu,~B.; Liu,~L.; Xie,~Z.; Ma,~Y. Magic-{Angle} {Stacking} and {Strong} {Intermolecular} $\pi$-$\pi$ {Interaction} in a {Perylene} {Bisimide} {Crystal}: {An} {Approach} for {Efficient} {Near}-{Infrared} ({NIR}) {Emission} and {High} {Electron} {Mobility}. \emph{The Journal of Physical Chemistry Letters} \textbf{2018}, \emph{9}, 596--600, Publisher: American Chemical Society\relax
\mciteBstWouldAddEndPuncttrue
\mciteSetBstMidEndSepPunct{\mcitedefaultmidpunct}
{\mcitedefaultendpunct}{\mcitedefaultseppunct}\relax
\EndOfBibitem
\bibitem[Huang \latin{et~al.}(2011)Huang, Barlow, and Marder]{huang_perylene-34910-tetracarboxylic_2011}
Huang,~C.; Barlow,~S.; Marder,~S.~R. Perylene-3,4,9,10-tetracarboxylic {Acid} {Diimides}: {Synthesis}, {Physical} {Properties}, and {Use} in {Organic} {Electronics}. \emph{The Journal of Organic Chemistry} \textbf{2011}, \emph{76}, 2386--2407, Publisher: American Chemical Society\relax
\mciteBstWouldAddEndPuncttrue
\mciteSetBstMidEndSepPunct{\mcitedefaultmidpunct}
{\mcitedefaultendpunct}{\mcitedefaultseppunct}\relax
\EndOfBibitem
\bibitem[May \latin{et~al.}(2011)May, Marcon, Hansen, Grozema, and Andrienko]{may_relationship_2011}
May,~F.; Marcon,~V.; Hansen,~M.~R.; Grozema,~F.; Andrienko,~D. Relationship between supramolecular assembly and charge-carrier mobility in perylenediimide derivatives: {The} impact of side chains. \emph{Journal of Materials Chemistry} \textbf{2011}, \emph{21}, 9538--9545, Publisher: The Royal Society of Chemistry\relax
\mciteBstWouldAddEndPuncttrue
\mciteSetBstMidEndSepPunct{\mcitedefaultmidpunct}
{\mcitedefaultendpunct}{\mcitedefaultseppunct}\relax
\EndOfBibitem
\bibitem[Mukazhanova \latin{et~al.}(2023)Mukazhanova, Negrin-Yuvero, Freixas, Tretiak, Fernandez-Alberti, and Sharifzadeh]{mukazhanova_impact_2023}
Mukazhanova,~A.; Negrin-Yuvero,~H.; Freixas,~V.~M.; Tretiak,~S.; Fernandez-Alberti,~S.; Sharifzadeh,~S. The impact of stacking and phonon environment on energy transfer in organic chromophores: computational insights. \emph{Journal of Materials Chemistry C} \textbf{2023}, \emph{11}, 5297--5306, Publisher: The Royal Society of Chemistry\relax
\mciteBstWouldAddEndPuncttrue
\mciteSetBstMidEndSepPunct{\mcitedefaultmidpunct}
{\mcitedefaultendpunct}{\mcitedefaultseppunct}\relax
\EndOfBibitem
\bibitem[Cullum and Willoughby(1981)Cullum, and Willoughby]{cullum_computing_1981}
Cullum,~J.; Willoughby,~R.~A. Computing eigenvalues of very large symmetric matrices—{An} implementation of a {Lanczos} algorithm with no reorthogonalization. \emph{Journal of Computational Physics} \textbf{1981}, \emph{44}, 329--358\relax
\mciteBstWouldAddEndPuncttrue
\mciteSetBstMidEndSepPunct{\mcitedefaultmidpunct}
{\mcitedefaultendpunct}{\mcitedefaultseppunct}\relax
\EndOfBibitem
\bibitem[Wilkinson(1988)]{wilkinson_algebraic_1988}
Wilkinson,~J.~H. \emph{The algebraic eigenvalue problem}; Oxford University Press, Inc.: USA, 1988\relax
\mciteBstWouldAddEndPuncttrue
\mciteSetBstMidEndSepPunct{\mcitedefaultmidpunct}
{\mcitedefaultendpunct}{\mcitedefaultseppunct}\relax
\EndOfBibitem
\bibitem[noa()]{noauthor_lanlpyseqm_nodate}
lanl/{PYSEQM}: an interface to semi-empirical quantum chemistry methods implemented with pytorch. \url{https://github.com/lanl/PYSEQM/}\relax
\mciteBstWouldAddEndPuncttrue
\mciteSetBstMidEndSepPunct{\mcitedefaultmidpunct}
{\mcitedefaultendpunct}{\mcitedefaultseppunct}\relax
\EndOfBibitem
\end{mcitethebibliography}

\end{document}